\documentclass[prl,aps,10pt,superscriptaddress,showpacs,reprint]{revtex4-1}
\usepackage[utf8]{inputenc}
\usepackage{amsmath,amsfonts,amsthm,amssymb}
\usepackage{graphicx}
\usepackage{hyperref}
\usepackage{graphics}
\usepackage{subfigure}
\usepackage{color}
\usepackage{dsfont}

\usepackage{braket}
\usepackage{setspace}
\usepackage{bm}
\usepackage{xcolor} % Only for "\comment"

\usepackage[normalem]{ulem} % strikeout: \sout

\newcommand{\D}{{\rm d}}
\newcommand{\CN}{{\cal N}}
\newcommand{\CO}{{\cal O}}
\def\E{{\mathbb E}}

\def\V{{\mathbb V}}

\newcommand{\CC}{{\cal C}}

\newcommand{\tr}{\operatorname{Tr}}
\newcommand{\Tr}{\tr}
\newcommand{\rd}{{\rm d}}
\newcommand{\density}{{g}}

\newcommand{\vecnotation}[1]{{\bm{#1}}}

\begin{document}
\title{How long does it take to obtain a physical density matrix?}

\author{Lukas~Knips}
\affiliation{Max-Planck-Institut f\"ur Quantenoptik, Hans-Kopfermann-Strasse 1, D-85748 Garching, Germany}
\affiliation{Department f\"ur Physik, Ludwig-Maximilians-Universit\"at, D-80797 M\"unchen, Germany}

\author{Christian~Schwemmer}
\affiliation{Max-Planck-Institut f\"ur Quantenoptik, Hans-Kopfermann-Strasse 1, D-85748 Garching, Germany}
\affiliation{Department f\"ur Physik, Ludwig-Maximilians-Universit\"at, D-80797 M\"unchen, Germany}

\author{Nico~Klein}
\affiliation{Max-Planck-Institut f\"ur Quantenoptik, Hans-Kopfermann-Strasse 1, D-85748 Garching, Germany}
\affiliation{Department f\"ur Physik, Ludwig-Maximilians-Universit\"at, D-80797 M\"unchen, Germany}

\author{Jonas~Reuter}
\affiliation{Bethe Center for Theoretical Physics, Universit\"at Bonn, D-53115 Bonn, Germany}

\author{G\'eza~T\'oth}
\affiliation{Department of Theoretical Physics, University of the Basque Country UPV/EHU, P.O. Box 644, E-48080 Bilbao, Spain}
\affiliation{IKERBASQUE, Basque Foundation for Science, E-48013 Bilbao, Spain}
\affiliation{Wigner Research Centre for Physics, Hungarian Academy of Sciences, P.O. Box 49, H-1525 Budapest, Hungary}

\author{Harald~Weinfurter}
\affiliation{Max-Planck-Institut f\"ur Quantenoptik, Hans-Kopfermann-Strasse 1, D-85748 Garching, Germany}
\affiliation{Department f\"ur Physik, Ludwig-Maximilians-Universit\"at, D-80797 M\"unchen, Germany}

\begin{abstract}
The statistical nature of measurements alone easily causes unphysical estimates in quantum state tomography.
We show that multinomial or Poissonian noise results in eigenvalue distributions converging to the Wigner semicircle distribution for already a modest number of qubits.
This enables to specify the number of measurements necessary to avoid unphysical solutions as well as a new approach to obtain physical estimates.
\end{abstract}
% 50-word abstract
% This letter clarifies the origin of statistical errors in Quantum State Tomography and introduces handy tools for state analysis. Based on our findings, one is able to determine the number of measurements required to obtain physical estimates or to identify systematic deviations like colored noise in the experimental setup. Our work considers the most relevant tomography methods, making it of general interest.

\maketitle

\setstretch{1.0}

\textit{Introduction.---}Quantum state tomography became the standard tool for fully determining unknown quantum states~\cite{ParisRehacek,EfficientTomo} %,TomographyPILike,CompressedSensing,MPS} 
and has been used for various systems like photons~\cite{Dicke63}, ions~\cite{IonTomography}, and cold gases~\cite{ColdGases}.
However, in experiments one often obtains unphysical estimates~\cite{Smithey}.
In principle, they can be mapped onto physical ones using numerical methods like the maximum likelihood estimation (ML) or least squares methods (LS)~\cite{Hradil, James}. 
Yet, the constraint of physicality results in biased estimation leading to distorted results and systematic errors of evaluted parameters~\cite{Bias,Sugiyama}.
Naturally, the question arises how unphysical estimates can be avoided.
Contrary to frequent folklore, experimental imperfections %faulty experimental equipment %bad experiments 
and misalignment are definitely not the only reasons~\cite{BlumeKohout}. 

Here, we demonstrate how statistical noise alone almost unavoidably causes unphysical estimates for multiqubit states. 
It is shown that for typical Poissonian or multinomial measurement statistics, the distribution of eigenvalues for a large number of qubits can be described by a Wigner semicircle distribution~\cite{Wigner,Mehta,HiaiPetz} with the width depending on the total number of measurements or observations like counts or clicks.
This now enables to specify how likely an unphysical solution is or, alternatively, to give a minimum number of measurements necessary to avoid unphysical solutions.
In turn, knowing the distribution of eigenvalues allows both to obtain a physical estimate from ``unphysical'' measurement data as well as to analyze possible misalignment or colored noise by hypothesis testing.

\textit{Linear Quantum State Estimation.---}The measurement scheme we first focus on is the so-called Pauli tomography scheme. 
%As measurement scheme we first focus on the so-called Pauli tomography scheme. 
There, for characterizing an $n$-qubit system, the individual qubits are projected on the eigenstates of all $3^n$ possible tensor products of local Pauli operators, $\sigma_1$, $\sigma_2$, and $\sigma_3$. 
%projective measurements of the individual qubits are performed on the eigenstates of all $3^n$ possible tensor products of local Pauli operators, $\sigma_1$, $\sigma_2$, and $\sigma_3$. 
For each of the $3^n$ measurement settings denoted by $\vecnotation{s}\in\{(s_1,s_2,\dots,s_n)|s_j\in\{1,2,3\},\; j=1,\dots,n\}$, one obtains $2^n$ different outcomes, denoted by $\vecnotation{r}\in\{(r_1,r_2,\dots,r_n)|r_j\in\{-1,1\},\; j=1,\dots,n\}$.  
This leads to a tomographically overcomplete set of $6^n$ relative frequencies $f_\vecnotation{r}^\vecnotation{s}=c_\vecnotation{r}^\vecnotation{s}/N_\vecnotation{s}$ with $c_\vecnotation{r}^\vecnotation{s}$ the number of results $\vecnotation{r}$ for settings $\vecnotation{s}$ and with total number of counts $N_\vecnotation{s}=\sum_\vecnotation{r} c_\vecnotation{r}^\vecnotation{s}$ of setting $\vecnotation{s}$. 
%For the linearly estimated state, one has to find the quantum state $\varrho_0$ whose probability to measure the outcome $r$ in basis $s$ equals the observed frequency $f_r^s$. 
% With the measurement operators $M_r^s$ one obtains a set of probabilities $P_{\varrho_0}^s(r)=\Tr(\varrho_0 M_r^s)$. 
% The linear estimate of the state is then given by linearly inverting these equations, i.e.,
We use the same total number of counts for all settings, $N_{\vecnotation{s}}=N \, \forall \, \vecnotation{s}$.

To determine the estimate $\tilde{\varrho}$ of the density matrix $\varrho$, the relation 
\begin{equation}
\varrho = \frac{1}{2^n} \sum_{\vecnotation{\mu}} T_{\vecnotation{\mu}} \sigma_{\vecnotation{\mu}}
\label{eq:statecorrelationtensor}
\end{equation}
can be used, where $T_{\vecnotation{\mu}}$ are the elements of the correlation tensor given by the expectation values of the respective settings $\vecnotation{\mu}$, $T_{\vecnotation{\mu}}=\langle\sigma_{\vecnotation{\mu}}\rangle$ ($\sigma_{\vecnotation{\mu}} = \sigma_{\mu_1} \otimes \sigma_{\mu_2} \otimes \cdots \otimes \sigma_{\mu_n}$, $\mu_{i}\in\{0,1,2,3\}$, and $\sigma_{0}$ denotes the identity matrix).
Estimates $\tilde{T}_{\vecnotation{\mu}}$ are obtained from the respective frequencies from which $\tilde{\varrho}$ results using Eq.~(\ref{eq:statecorrelationtensor})~\cite{KieselPhd}.
In practice, due to the finite number of measurements and the resulting statistical noise, the $\tilde{T}_{\vecnotation{\mu}}$ and consequently the elements of $\tilde{\varrho}$ are random variables depending on both the statistics and the observed state.
Note, for overcomplete tomography, e.g., measuring all $3^n$ possible settings, the different elements of $\tilde{\varrho}$ do not have all the same uncertainty.
For example, due to the inherent redundancy, if one of the $\mu_i$, say, $\mu_k=0$, all $3\times2^{n}$ frequencies $f_\vecnotation{r}^\vecnotation{s}$ obtained from the three settings $\vecnotation{s}=(s_1,\dots,s_{k-1},s_k,s_{k+1},\dots,s_{n})$ with $s_k=\{1,2,3\}$ can be used to calculate $\tilde{T}_{\vecnotation{\mu}}$, thus reducing its variance.
These correlations are called \textit{non-full correlations}, while those with $\mu_k\neq0$ for all $k$ are \textit{full correlations}.
For more details, see SM~1~\cite{Supplement}.
%Effectively, all $6$ frequencies $f_r^s$ of the three settings $s=(s_1,\dots,s_{k-1},s_k,s_{k+1},\dots,s_{n})$ with $s_k=\{1,2,3\}$ can be used.

\textit{Eigenvalue distribution for many qubits.---}%Before we will generalize our study to a more general class of states, we first restrict ourselves to the case of the completely mixed state. 
To understand how finite statistics influences the distribution of eigenvalues, the completely mixed state $\varrho_{\textrm{wn}}$ (white noise) is chosen as starting point. 
This state is characterized by $T_{\vecnotation{\mu}}=0$ for all $\vecnotation{\mu}$ except $T_{0,\dots,0}=1$, i.e.,
\begin{equation}
\varrho_{\textrm{wn}} = \frac{1}{2^n} \sigma_{0,0,\dots,0} = \frac{1}{2^n} \mathds{1}
\end{equation}
with $2^n$ degenerate eigenvalues $\lambda_i=1/2^n$. 
% Using $\varrho_{\textrm{cm}}$ for QST, we expect all non-trivial correlations to vanish, i.e., $\E\left(T_{\vecnotation{\mu}}\right)=0$ for all $\vecnotation{\mu}\neq(0,\dots,0)$. 
Statistical fluctuations of the measurement result $c_{\vecnotation{r}}^{\vecnotation{s}}$ and of the resulting $\tilde{T}_{\vecnotation{\mu}}$ lift the degeneracy and consequently cause a wide distribution of observed eigenvalues.
With Poissonian or multinomial noise for $c_{\vecnotation{r}}^{\vecnotation{s}}$, the elements of the estimated density matrix $\tilde{\varrho}$ can be approximated by a Gaussian distribution.
% Due to statistical fluctuations, experimentally obtained correlations will differ from the expectation values. 
%We prove in the Supplementary Material (SM)~\cite{Supplement} that the spectral distribution of such matrices converges in the limit of large matrices, i.e., of many qubits, to the Wigner semicircle distribution~\cite{Mehta,Wigner}. 
In the limit of large matrices, the eigenvalues are distributed according to a Wigner semicircle distribution~\cite{Wigner,Mehta,HiaiPetz} %$f_{c,R}(x)=\frac{2}{\pi R^2}\sqrt{\left(x-c\right)^2-R^2}$ 
(see SM~2~\cite{Supplement} for a proof for the used overcomplete Pauli tomography).
\begin{figure}
\includegraphics[width=0.48\textwidth]{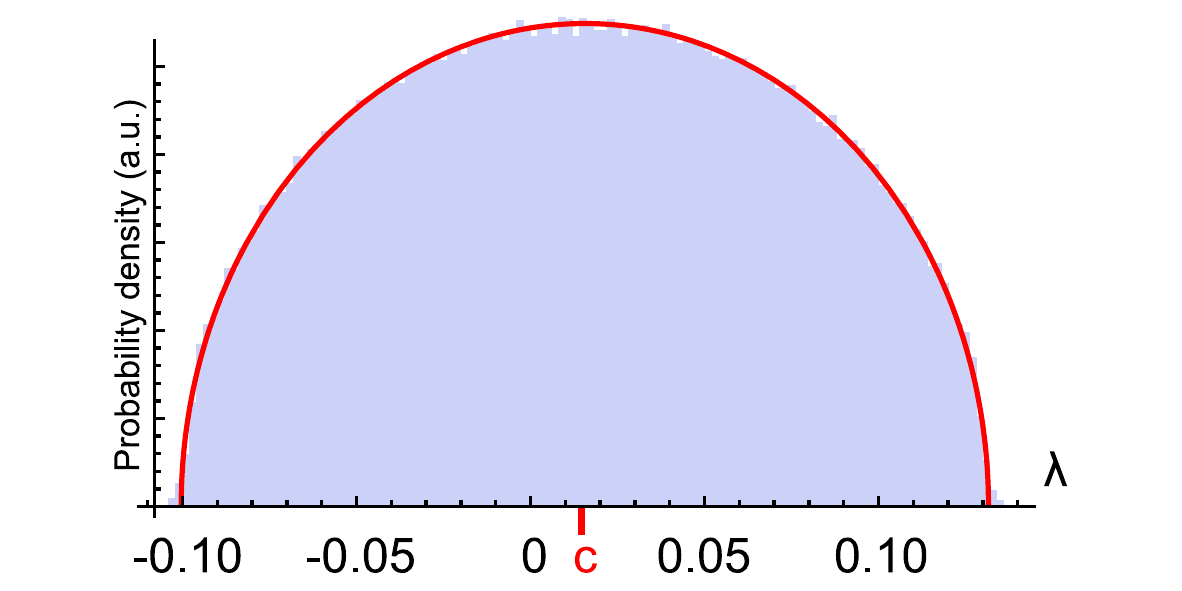}
\caption{The relative occurences of eigenvalues for the $n=6$ qubit completely mixed state. 
$10\,000$ QST and linear state estimations have been performed on simulated data with $N=100$ expected counts per measurement basis. 
The resulting eigenvalues are shown in the blue histogram. 
The semicircle (red) is centered around $c=2^{-6}$ with a radius of about $R=0.116$. 
%Note that the ellipticity of the red line depends on the normalization and is chosen to match the binning. 
Center and radius are given by Eq.~(\ref{eq:center}) and Eq.~(\ref{eq:effectiveRadius}), respectively.}
\label{fig:wn_n6_N100}
\end{figure}
Yet, already for a modest number of $n=6$ qubits, we observe an excellent agreement as shown in Fig.~\ref{fig:wn_n6_N100}.
%The Wigner semicircle distribution can be used already for a small number of qubits as an approximation, as Fig.~\ref{fig:wn_n6_N100} shows for the completely mixed state with $n=6$ qubits.
%To characterize this distribution, we have to estimate its center and radius. 
%Obviously, the semicircle's center $c$ is determined by the expected mean value of the eigenvalues $\lambda_i$, i.e.,
The center $c$ of the distribution is given as the mean value of the eigenvalues $\lambda_i$ and is, due to normalization, equal to
\begin{equation}
c = \frac{1}{2^n}\E\left[\sum_{i=1}^{2^n}\lambda_i\right]=\frac{1}{2^n},
\label{eq:center}
\end{equation}
%while the radius $R$ depends on the variance of the correlation tensor elements $\Delta^2_{T_{\vecnotation{\mu}}}=\E\left( T_{\vecnotation{\mu}}^2 \right)-\E\left( T_{\vecnotation{\mu}} \right)^2$ via the relation (see SM~2~\cite{Supplement})
while the radius $R$ depends on the second moment of the correlation tensor elements $\E\left( \tilde{T}_{\vecnotation{\mu}}^2 \right)$ via the relation (see SM~2~\cite{Supplement})
\begin{equation}
\left(\frac{R}{2}\right)^{2}=\frac{1}{4^n} \sum_{\vecnotation{\mu}} \E\left( \tilde{T}_{\vecnotation{\mu}}^2 \right).
\label{eq:momentsk1}
\end{equation}
For the completely mixed state and multinomial noise, the probability density $g(\tilde{T}_{\vecnotation{\mu}})$ with $N$ counts can be approximated as $g(\tilde{T}_{\vecnotation{\mu}})\approx\sqrt{N/\left(2 \pi\right)}\exp\left(-\tilde{T}_{\vecnotation{\mu}}^2 N / 2\right)$, see SM~1~\cite{Supplement}.
%$\density(T_{\vecnotation{\mu}})=\sqrt{N/\left(2 \pi\right)}\exp\left(-T_{\vecnotation{\mu}}^2 N / 2\right)$. 
%Thus, we can evaluate the expectation value for full correlations to be
Thus, we can evaluate the second moment of full correlations to be
% \begin{equation}
$\E\left(\tilde{T}_{\vecnotation{\mu}}^2\right) = \int_{-\infty}^{\infty} \rd \tilde{T}_{\vecnotation{\mu}} g(\tilde{T}_{\vecnotation{\mu}}) \tilde{T}_{\vecnotation{\mu}}^2 \approx \frac{1}{N}$. 
% \label{eq:statisticsFullCorr}
% \end{equation}
For non-full correlations we obtain
% \begin{equation}
$\E\left(\tilde{T}_{\vecnotation{\mu}}^2\right) \approx \frac{1}{3^{j\left(\vecnotation{\mu}\right)} N}$
% \label{eq:statisticsNonFullCorr}
% \end{equation}
with $j\left(\vecnotation{\mu}\right)$ denoting the number of $\sigma_0$ operators in $\sigma_\vecnotation{\mu}$. %of local measurements of $\sigma_0$. %, i.e. $|\{\mu_i:\,\mu_i=0\}|=j$. 
Using these expressions for the evaluation of the right hand side of Eq.~\ref{eq:momentsk1}, we obtain
%Taking the respective statistics of all occuring correlations into account, i.e., using the above expression for all correlations $\vecnotation{\mu}$, we obtain
\begin{equation}
\frac{1}{4^n} \sum_{\vecnotation{\mu}} \E\left(\tilde{T}_{\vecnotation{\mu}}^2\right) = \frac{1}{4^n} \sum_{j=0}^{n-1} 3^{n-j} \binom{n}{j} \frac{1}{3^j N} = \frac{10^n - 1}{12^n N}.
\end{equation}
Hence, the radius $R$ of the semicircular approximation of the $n$-qubit completely mixed state is given by
\begin{equation}
R = 
%2 \sqrt{\frac{1}{4^n} \sum_{\vecnotation{\mu}} \E\left(T_{\vecnotation{\mu}}^2\right)} = 
2 \sqrt{\frac{10^n - 1}{12^n}}\frac{1}{\sqrt{N}}\approx2\left(\frac{5}{6}\right)^{\frac{n}{2}}\frac{1}{\sqrt{N}}.
\label{eq:effectiveRadius}
\end{equation}
Note, the radius of the semicircle approximation decreases with the statistics $N$ as $\sqrt{1/N}$. 
For the case of $n=6$ qubits, which are represented by $64\times64$ matrices, and $N=100$ as typically used in recent multiqubit experiments, we can now estimate the effective radius to be $R=0.116$. 
The red semicircle in Fig.~\ref{fig:wn_n6_N100} is in clear agreement with the distribution of eigenvalues obtained from simulated quantum state tomographies. 
Obviously, if the radius $R$ is larger than the center $c$, there is a non-vanishing probability for negative eigenvalues.
% Obviously, for a radius $R$ ($R<c$), there is a non-vanishing probability for negative eigenvalues.

\textit{Eigenvalue distribution for few qubits.---}
The Wigner semicircle describes the spectral distribution in the limit of large matrices, i.e., of many qubits~\cite{Mehta,Wigner}.
% The proof of the semicircular spectral distribution as done, e.g., in ~\cite{Supplement,Mehta,Wigner} is based on the limit of large matrices, i.e., of many qubits. 
Let us thus analyze also the case of states with few qubits.
% Here, we will analytically study the case of a single qubit and compare the spectral distributions of states with few qubits. 
% Since the eigenvalues of $n=1$ qubit states can be easily expressed in an analytic manner, the spectral density can be directly evaluated to be~\cite{Supplement}
For a single qubit, the eigenvalues and the spectral density can be determined analytically, 
\begin{equation}
g(\lambda) \propto \exp\left[-\frac{\left(1-2\lambda\right)^2 N}{2}\right] \left(1-2\lambda\right)^2,
\label{eq:analytical_solution_1qubit_wn}
\end{equation}
see SM~4~\cite{Supplement}.
\begin{figure}
\includegraphics[width=0.48\textwidth]{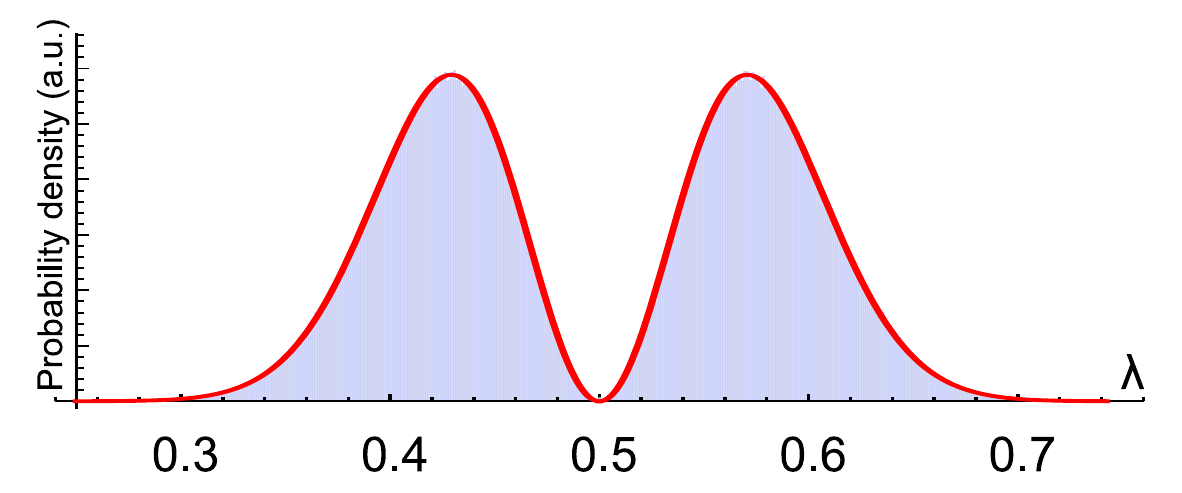}
\caption{The spectral probability distribution of the single qubit completely mixed state. 
QST was simulated $1\,000\,000$ times, each with $N=100$ events per measurement setting. (blue area) Histogram of the eigenvalues with high-resolution bins. (red line) Curve proportional to the expected distribution, cf. Eq.~(\ref{eq:analytical_solution_1qubit_wn}).}
\label{fig:wn_n1_N100}
\end{figure}
Fig.~\ref{fig:wn_n1_N100} shows the simulated spectral probability of the $n=1$ qubit completely mixed state together with the probability density, Eq.~(\ref{eq:analytical_solution_1qubit_wn}).
Evidently, it is unlikely to obtain the correct result with both eigenvalues at $1/2$ since, due to $\lambda_{1,2} = \frac{1}{2} \left(1 \pm \sqrt{T_1^2+T_2^2+T_3^2}\right)$, all correlations $T_1$, $T_2$ and $T_3$ have to vanish for this case. 
In particular, if $N$ is odd, it is even impossible. 
Furthermore, we see an exponential damping towards the boundaries of the distribution. 
However, for resonable number of measurements, the probability of negative eigenvalues is negligible.

\begin{figure}
\includegraphics[width=0.48\textwidth]{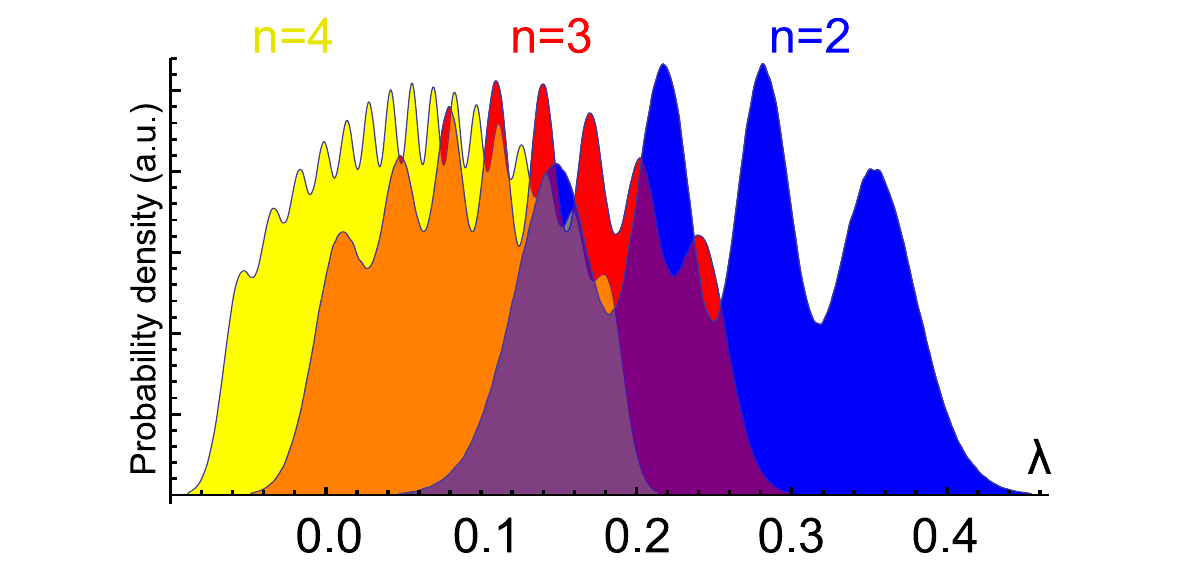}
\caption{Smoothed histogram (binning $0.001$) of the distribution of eigenvalues of simulated QSTs of the completely mixed state.
Results for increasing qubit numbers are shown from right to left.
(blue) Results for $n=2$ qubits for $N=100$, displaying four peaks.
The center of the distribution is given by Eq.~(\ref{eq:center}). 
The two-qubit state remains in almost all cases physical, i.e., all eigenvalues can be interpreted as probabilities.
(red) For $n=3$ qubits, the distribution is shifted such that many eigenvalues are below zero, indicating a high probability for unphysical matrices. 
(yellow) For $n=4$ qubits, all states resulting from simulated QST lack physicality. 
Additionally, one perceives the increased damping towards the boundaries of the distribution for $n=4$ compared to $n=3$ and $n=2$.}
\label{fig:wn_n2_n3_n4_N100}
\end{figure}

While the spectral distribution in the case of a single qubit obviously differs from a semicircular behavior, we learn about the transition towards the limiting case by comparing the single qubit case (Fig.~\ref{fig:wn_n1_N100}) with the distributions of the completely mixed states for $n=2$, $3$, $4$ qubits, respectively. 
Fig.~\ref{fig:wn_n2_n3_n4_N100} indicates that by increasing the number of qubits the distribution develops a comb structure. 
The center of the distribution shifts, according to Eq.~(\ref{eq:center}), closer to $0$. 
As the probability density then extends to negative values, it becomes increasingly unlikely, even for the completely mixed state, to obtain physical results when performing quantum state tomography for a limited number of measurements (here $N=100$). 
In the case of $n=2$ qubits, shown by the blue histogram in Fig.~\ref{fig:wn_n2_n3_n4_N100}, a fraction of only $6\times10^{-6}$ of the simulated states is unphysical. 
In contrast, for $n=3$ qubits, already $32\%$ of the estimated states lack physicality, while for $n=4$ \textit{all} estimates were unphysical (with $1\,000\,000$ simulated QST in each case). 
The distribution becomes increasingly damped towards the boundaries, leading to the semicircular distribution in the limit of many qubits.

\textit{Different tomography schemes.---}
The overcomplete tomography is one of several sampling strategies. 
The resulting structure of variances of the density matrix elements of the linear estimate leads to the Wigner semicircle distribution. 
For alternative tomography schemes, e.g.~\cite{OptimalTomography}, one directly obtains the individual matrix elements and thus can taylor the respective variances to be the same for all elements. 
From random matrix theory it is known that equal Gaussian distribution of the matrix elements also leads to a semicircle distribution for the eigenvalues, in this case with a radius of $R = 2\tau$ with the variance of the matrix elements $\tau^2=1/N$ ~\cite{Bai} (see also SM~2~\cite{Supplement}). %\textcolor{red}{$R = 1 2 3$}~\cite{Bai}.
Similarly, a symmetric sampling procedure~\cite{SICQST} will yield again such a distribution~\cite{QSTRandomTheory}.  

The situation changes completely for the most frequently used complete tomography scheme~\cite{James}. 
%[Although the individual matrix elements again have approximately Gaussian distribution, the symmetry of the sampled correlation terms is lost and does not allow anymore to form the moments of the semicircle function.] 
%\textcolor{red}{Explaining sentence is missing. }
There, the projectors are orientated asymmetrically, leading to different variances of the correlations and, thus, to strongly differing variances of the matrix elements other than for the overcomplete scheme.
This inhomogeneity of the variances results in a probability density function of an exponential distribution with a width significantly larger than the corresponding semicircle radius for an equal total number of measurements, see SM~6~\cite{Supplement}. 
Due to the narrower distribution and the resulting higher probability to obtain a physical estimate it is thus definitely more favourable to choose a symmetric tomography scheme or the overcomplete Pauli scheme.

\textit{Required number of measurements.---}The knowledge of the eigenvalue spectrum of white noise now enables to estimate the necessary measurement statistics required to obtain a physical density matrix~\cite{PathologicalCases}.
An experimentally relevant state is a pure or low rank state of rank $r$ mixed with white noise of the form
%The theory of random matrices does not only apply for the completely mixed state, but instead can be used for an approximate description of other states as well. 
%While this survey can also be generalized for other classes, we study the case of states of the experimentally interesting form
\begin{equation}
%\varrho_{q,|\psi\rangle} = q |\psi\rangle\langle\psi| + (1-q) \varrho_{\rm cm}.
{\varrho}= q_r \varrho_{r} + \left(1-q_r\right)\varrho_{\mathrm{wn}}.
\label{eq:statedefinition}
\end{equation}
%where a pure state $|\psi\rangle$ is mixed with $\varrho_{\rm cm}$. 
%Up to now, we considered only $q=0$, but the estimated radius of Eq.~(\ref{eq:effectiveRadius}) describes the probability distribution of the degenerate eigenvalues also for this general case. 
%Nevertheless, the center of the semicircle distribution is now shifted to
%For the mixing parameter $q$, the probability of observing $\ket{\psi}\bra{\psi}$ is $q+(1-q)/2^n$, which consequently will be one of the eigenvalues of $\varrho_{q,\ket{\psi}}$. 
For a pure state mixed with white noise ($r=1$, $\varrho_r=\ket{\psi}\bra{\psi}$), the probability of observing $\ket{\psi}$ is $q+(1-q)/2^n$, which consequently will be one of the eigenvalues of ${\varrho}$. 
The other eigenvalues are degenerate and will be distributed, for finite statistics, again according to a semicircle distribution, but now centered significantly closer to $0$,
\begin{equation}
c_{q,r} = \frac{1-q}{2^n - r}.% \Rightarrow c_{q,1} = \frac{1-q}{2^n - 1}.
\label{eq:center_generalized}
\end{equation}

\begin{figure}
\includegraphics[width=0.48\textwidth]{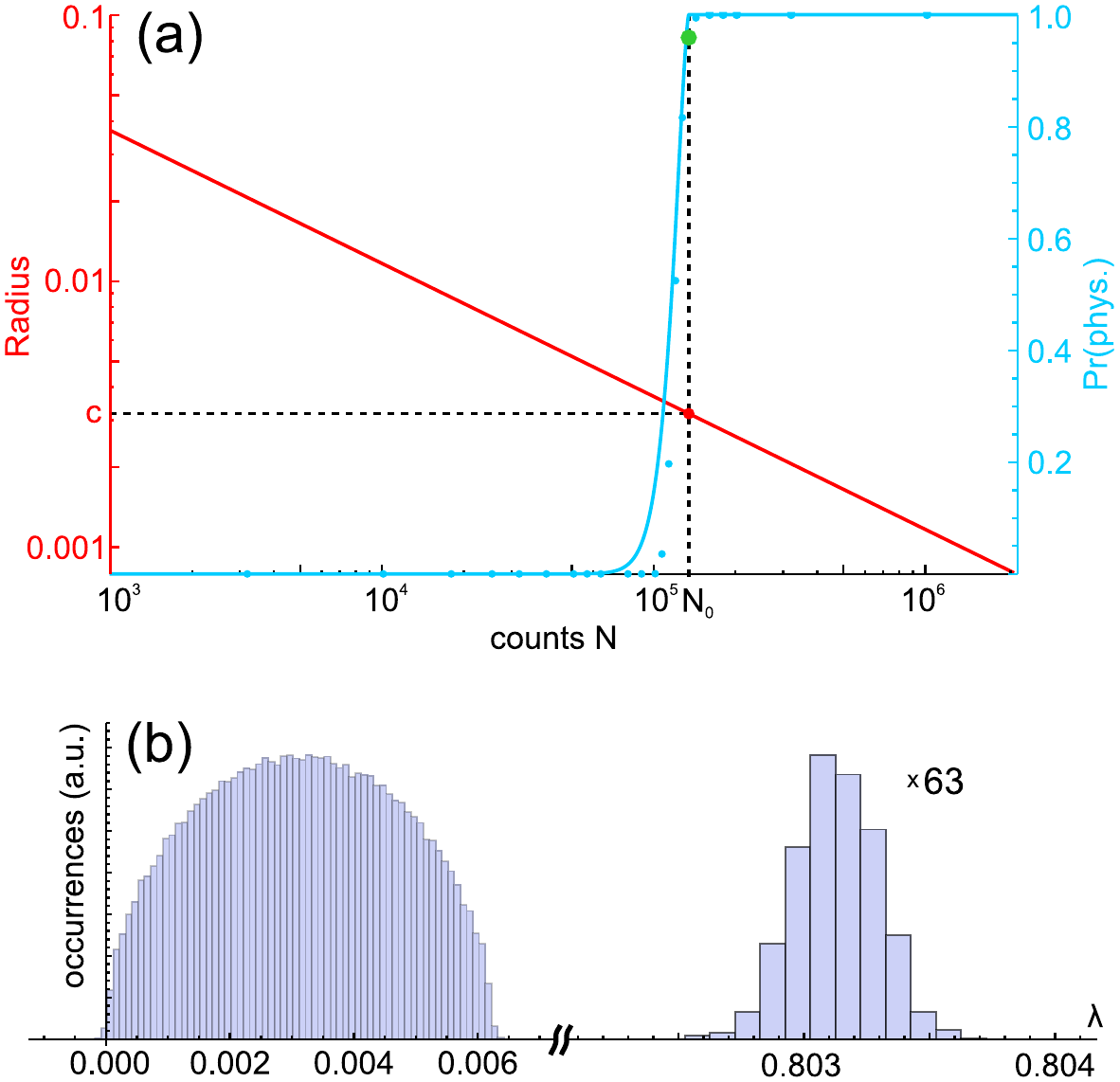}
\caption{
%Determining the number of events $N_0$ needed for a physical density matrix. 
Verifying our estimate $N_0$ for the number of events needed for a physical state, given in Eq.~(\ref{eq:minimalNumberOfCounts}). 
(a) (red line, left $y$-axis) The radius scales with $1/\sqrt{N}$ with the number of events per basis setting $N$ according to Eq.~(\ref{eq:effectiveRadius}), shown here for $n=6$ qubits. 
For a given admixture of white noise $1-q$, $N_0$ is obtained by equating the radius with center given in Eq.~(\ref{eq:center_generalized}) as a function of $q$. 
For $q=0.8$, we obtain $c=0.0032$ and $N_0=132\,921$. 
(blue dots) The ratio of physical states for simulated QST for the given noise parameter. %The large green circle corresponds to the number of counts \eqref{eq:minimalNumberOfCounts} needed to expect a physical state, where the data of (b) are taken. 
(blue line) Theoretical results based on a perfect Wigner semicircle and assuming independent distributions for the eigenvalues, see SM~5~\cite{Supplement}.  
%Approximative theory derived from the Wigner semicircle, see SM~5~\cite{Supplement}. 
(b) Histogram of eigenvalues for a GHZ state with admixed white noise for $N_0$ events per setting [green circle in (a)]. 
%b) Distribution of eigenvalues of the admixed white noise and
The largest eigenvalues of each simulation correspond (approximately) to the GHZ state.
Note that each simulated state results in $63$ eigenvalues distributed within the semicircle and a single eigenvalue around $0.8+0.2/64\approx0.803$.
% (a) Eigenvalues corresponding to the admixed white noise. Eigenvalues below $0$, indicating non-physicality, are marked red.
% (b) Distribution of the 
% eigenvalues of the GHZ state.
% Note that each state corresponds to $63$ eigenvalues in figure (a) and a single eigenvalue in figure (b).
%For details, see text. 
%Thus, the probability $\operatorname{Prob}\left(\varrho\nleq0\right)$ (blue dashed lines) to obtain an unphysical estimate for $\varrho$ for almost pure states ($p$ close to $1$) is still large for high $N$.
}
\label{fig:radiusWithCounts_n6}
\end{figure}

% \begin{figure}
% \includegraphics[width=0.42\textwidth]{pictures/ghz_n6_F80_N128800_2.pdf}
% \caption{(color online) Numerical test of the formula \eqref{eq:minimalNumberOfCounts} giving the minimal number of measurements to obtain a physical density matrix for an $n=6$ qubit GHZ state. The state is mixed with white noise as given in (\ref{eq:statedefinition}), where $q$ is given in the text.
% (a) Eigenvalues corresponding to the admixed white noise. Eigenvalues below $0$, indicating non-physicality, are marked red.
% (b) Distribution of the 
% eigenvalues of the GHZ state.
% Note that each state corresponds to $63$ eigenvalues in figure (a) and a single eigenvalue in figure (b).
% }
% \label{fig:ghz_n6_F80_N128800_2}
% \end{figure}

For a low rank $r$ state admixed with white noise as for states of the form (\ref{eq:statedefinition}) with $r=1$, the number of eigenvalues in the support of the semicircle is changed ($2^n\rightarrow2^n-r$), see SM~7~\cite{Supplement}. 
This also leads to a reduction of the radius by a factor of $\sqrt{\left(2^n-r\right)/2^{n}}$, i.e.,
\begin{equation}
R \rightarrow R_r = R \cdot \sqrt{1-r\cdot2^{-n}}.
\end{equation}
%which will be neglected for low rank states with $r\ll2^n$.
% For rank $r=1$ states (admixed with white noise) the radius can be assumed to be approximately unchanged.
For pure states admixed with white noise the radius can be assumed to be approximately unchanged.
%The change in the number of eigenvalues in the support of the semicircle ($2^n\rightarrow2^n-1$) also leads to a modification of the radius by a factor of $\left(2^n-1\right)/2^n$.
%In the case of a low rank state admixed with white noise as for states of the form (\ref{eq:statedefinition}), this modification can be neglected and the radius can be assumed to be approximately unchanged, see SM~7~\cite{Supplement}
We can thus use the radius (\ref{eq:effectiveRadius}) and center (\ref{eq:center_generalized}) of the semicircle to give an estimate of the necessary amount of measurement events such that the data directly result in a physical density matrix. 
% Thus, we assess the amount of necessary events per measurement basis to be
A physical solution is expected for $R\leq c$ and thus for
\begin{equation}
%N\geq N_0=\frac{4 (10^n - 1)}{12^n}\left(\frac{2^n-1}{1-q}\right)^2.%\approx4\left(\frac{10}{3}\right)^n\left(\frac{1}{1-p}\right)^2.
N\geq N_0=4\left(\frac{5}{6}\right)^{n}\left(\frac{2^n-1}{1-q}\right)^2.%\approx4\left(\frac{10}{3}\right)^n\left(\frac{1}{1-p}\right)^2.
\label{eq:minimalNumberOfCounts}
\end{equation}
% such that for $N\gtrsim N_0$ the linearly estimated quantum state obeys the physicality constraints. 
% Fig.~\ref{fig:ghz_n6_F80_N128800_2} shows the distribution of the eigenvalues of an $n=6$ qubit Greenberger-Horne-Zeilinger (GHZ) state $|\mathrm{GHZ}\rangle=\frac{1}{\sqrt{2}}(|000000\rangle+|111111\rangle)$ with admixed white noise according to the noise model in Eq.~(\ref{eq:statedefinition}) for $q=0.8$, leading theoretically to a fidelity $F_{\rm GHZ}(\varrho_{q=0.8})=80.3125\%$. 
% We find that for about $N_0=132\,921$, radius and center of the Wigner semicircle approximation are equal ($c_{0.8}=R=0.0032$).
Fig.~\ref{fig:radiusWithCounts_n6} (a) gives a graphical access to this condition.
The red line depicts the dependence of $R$ on $N$ according to Eq.~(\ref{eq:effectiveRadius}) for $n=6$ qubits.
In order to find the probability for physical state estimates, for each $N$, more than $2\,500$ tomography data sets were simulated for a GHZ state admixed with white noise [see Fig.~\ref{fig:radiusWithCounts_n6} (b)].
%For $q=0.8$, we obtain $c_{0.8}=0.0032$ and $N_0=132\,921$.
%\footnote{The values used for our calculations were $q=0.8$ and $R=64/19845\approx0.0032.$}.
%For $N=\left\lfloor{N_0}\right\rfloor$ about $95.9\%$ of the %$2\,500$ 
For $N$ chosen to be approximately $N_0$, about $95.9\%$ of the %$2\,500$ 
simulated states were physical, where the deviation can be explained by the tails of the distribution in the case of a finite $n=6$ (see also Fig.~\ref{fig:wn_n6_N100}). %being in excellent agreement with our approximations. 
Furthermore, one directly recognizes that the denominator in Eq.~(\ref{eq:minimalNumberOfCounts}) diverges for $q\rightarrow1$.
This underlines the fact that it is practically impossible to observe a pure (or low rank) multiqubit state ($q=1$) without obtaining negative eigenvalues.

\textit{Verifying physical estimates.---}%estimation.---}
Given the problem of conventional QST schemes, resulting in biased state estimation~\cite{Bias}, one can now use our findings to devise a new strategy to obtain a physical state estimate from a given data set.
%As it is the experimentally most relevant case, we make the assumption that the physical state is the mixture of a low rank state $\varrho_{r}$ ($r$ the rank of $\varrho_r$) and white noise and try to identify $\varrho_{r}$, i.e., 
%\begin{equation}
%\hat{\varrho}= q_r \varrho_{r} + \left(1-q_r\right)\varrho_{\mathrm{cm}}.
%\end{equation}
As it is the experimentally most relevant case, we make the assumption that the physical state is the mixture of a low rank state $\varrho_{r}$ and white noise as given in Eq.~(\ref{eq:statedefinition}) and try to identify $\varrho_{r}$.
Evidently, for a noisy low rank state sampled with finite statistics the spectrum will consist of $2^n-r$ eigenvalues scattered within an interval of width 
\begin{equation}
%w_r=2\cdot R_r= 2\cdot2\left(\frac{5}{6}\right)^{\frac{n}{2}}\frac{1}{\sqrt{N}}\sqrt{1-r\cdot2^{-n}}
w_r=2\cdot R_r= 2\cdot2\left(\frac{5}{6}\right)^{\frac{n}{2}}\sqrt{\frac{1-r\cdot2^{-n}}{N}}
\end{equation}
%$2R$ 
and $r$ eigenvalues outside.
The main question is how many of the eigenvalues correspond to the low rank state $\varrho_{r}$. 
We propose to use the ansatz that $\varrho_{r}$ is composed of those eigenvalues larger than $\min_{i} \lambda_i + w_r$ and to utilize a hypothesis test for its validation. 
%This can be validated using a hypothesis test.

To illustrate the new scheme, we analyze the experimental measurement data of an overcomplete quantum state tomography of a $6$ qubit Dicke state obtained in a $6$ photon experiment~\cite{Dicke63}. 
%Comparing the anticipated distribution with experimental measurement results, one is now able to identify systematical deviations of the prepared state. 
%With this, one is now able to find an ansatz to estimate a physical state out of measurement data based on the assumption of a low rank state mixed with white noise as given in Eq.~(\ref{eq:statedefinition}).
%To validate our results, we use methods of hypothesis testing.
%We investigate the experimental measurement data of an overcomplete quantum state tomography of a $6$ qubit Dicke state with $3$ excitations~\cite{Dicke63}. 
%We compare the eigenvalues of the linearly reconstructed matrix with the anticipated distribution. 
On average, $N=230$ projection measurements per setting were performed~\cite{Dicke63}, leading to $R_r \approx 0.07632\cdot\sqrt{1-r\cdot2^{-n}}$. 
%The center $c$ of the semicircular distribution is given by the amount of white noise and the number of degenerate eigenvalues. 
The $64$ eigenvalues of the state estimate are found to be $\{\lambda_i\}=\{-0.06368$, $-0.06223$, $\dots$, $0.06371$, $0.07171$, $0.14949$, $0.21595$, $0.61024\}$.
%\textcolor{red}{Now it is a fix point problem! }
For $r=0$ (assuming only white noise), the radius is $R_{r=0}=0.07632$, resulting in $3$ eigenvalues significantly larger than $\lambda_1 + w\left(0\right) = 0.08832$ (for $\lambda_1<\dots<\lambda_i<\lambda_{i+1}<\dots<\lambda_{64}$).
Any hypothesis test, e.g., the Anderson-Darling test~\cite{AndersonDarling,Mood}, which is based on a distance between the cumulative distribution function of the Wigner semicircle and the empirical distribution function (see also SM~8~\cite{Supplement}), clearly rejects this assumption.
Since $R_r$ changes only slightly with $r$, also the cases $r=1$ and $r=2$ have to be refused.
%Consequently, this assumption has the be rejected.
%For $r=1$ and $r=2$, one again finds $3$ eigenvalues outside of the support of the assumed semicircle, i.e., larger than $\lambda_1 +w\left(1\right) = 0.08777$ and $\lambda_1 + w\left(2\right) = 0.08656$.
%The spectrum is hence not in agreement with the assumption that the experiment produces a pure or a rank $r=2$ state, respectively, admixed with white noise, since $\lambda_{62}$ could not be explained by statistical noise only.
%%One directly sees that there are $3$ eigenvalues larger than $\lambda_1 + 2R = 0.08832$ (for $\lambda_1<\dots<\lambda_i<\lambda_{i+1}<\dots<\lambda_{64}$).
% Thus, the rank of $\varrho_{r}$ has to be set to $r=3$ with the center obtained as the average of the smallest $61$ eigenvalues or, respectively, by $c = \E\left(\{\lambda_i|i\leq61\}\right)=(1-\lambda_{64}-\lambda_{63}-\lambda_{62})/61 \approx 4\times 10^{-4}$. 
Setting the rank of $\varrho_{r}$ to $r=3$, the center is obtained as the average of the smallest $61$ eigenvalues or, respectively, by $c = \E\left(\{\lambda_i|i\leq61\}\right)=(1-\lambda_{64}-\lambda_{63}-\lambda_{62})/61 \approx 4\times 10^{-4}$. 
%The Anderson-Darling hypothesis test returns a $P$-value based on the distance $D\left(\lambda\right)={\left[\tilde{F}\left(\lambda\right)-F\left(\lambda\right)\right]^2}/\left[{F\left(\lambda\right)\left(1-F\left(\lambda\right)\right)}\right]$ with the cumulative distribution function $F\left(\lambda\right)$ of the Wigner semicircle and the empirical distribution function $\tilde{F}\left(\lambda\right)$ which is to test.
% The Anderson-Darling hypothesis test returns a $P$-value based on a distance of the cumulative distribution function of the Wigner semicircle and the empirical distribution function which is to test.
Applying the Anderson-Darling hypothesis test to our data results in a $P$-value of about $1-8.2\cdot10^{-7}$ which is well above the prechosen significance level (typically $0.05$), thus indicating that the data can indeed be explained by a low rank state of rank $r = 3$ with admixed white noise.

\textit{Conclusions.---}
For a finite number of measurements, the eigenvalues of the completely mixed state are distributed according to the Wigner semicircle in the limit of many qubits.
But also for a modest number of already $6$ qubits this distribution is already an excellent approximation.
Our findings enable one to determine the number of measurements required to obtain a physical solution.
Remarkably, the frequently used complete sampling strategy results in a different, much wider distribution and thus more often in unphysical estimates.
As the required number of measurements depends exponentially on the number of qubits, the goal to obtain a physical estimate $\tilde{\varrho}$ for a larger number of qubits is an illusion.
The knowledge of the possible structure of $\tilde{\varrho}$ allows a new ansatz for discriminating statistical effects from systematic ones or, vice versa, for analysing colored noise contributions.
Performing hypothesis tests gives a firm basis for the determination of a physical density matrix and thereby a new level of confidence in the tomography of multiqubit states.
%We proved that eigenvalues of the completely mixed state are distributed according to the Wigner semicircle in the limit of many qubits. 
%Based on simulated quantum state tomographies, we have shown that the semicircular distribution is already an appropriate approximation for only a few qubits. 
%This distribution is not restricted to the completely mixed state, but can also be used to gain information about experimentally prepared states. 
%Fortunately, it is even possible to distinguish statistical and systematic effects in measured quantum states, as has been shown with experimental data. 

\singlespacing

\textit{Acknowledgments.---}We like to thank O. G\"uhne, M. Kleinmann, T. Moroder for stimulating discussions. 
This work has been supported by the EU (QWAD, ERC AdG QOLAPS, ERC StG GEDENTQOPT),
by the MINECO (Project No. FIS2012-36673-C03-03), 
the Basque Government (Project No. IT4720-10), 
the UPV/EHU Program No. UFI 11/55, the OTKA (Contract No. K83858),
and the excellence cluster Nano-Initiative Munich (NIM). 
The work of JR is partially supported by the Bonn-Cologne Graduate school BCGS.
LK and CS acknowledge support by the Elite Network of Bavaria (QCCC and ExQM).

\cleardoublepage

% \onecolumngrid

\section*{\large Supplemental Material}

\section{SM~1: Gaussian approximation of correlations}
A correlation can be calculated by the normalized difference of the events contributing positively (indicating a \textit{correlation}), $c_{\vecnotation{\mu},\uparrow}$, and those that indicate an \textit{anticorrelation}, $c_{\vecnotation{\mu},\downarrow}$, i.e.,
\begin{equation}
\tilde T_{\vecnotation{\mu}} = \frac{c_{\vecnotation{\mu},\uparrow}-c_{\vecnotation{\mu},\downarrow}}{c_{\vecnotation{\mu},\uparrow}+c_{\vecnotation{\mu},\downarrow}}.
\label{eq:definition_correlation}
\end{equation}
Expecting uncorrelated results, i.e., $\E\left(\tilde T_{\vecnotation{\mu}}\right)=0$, leads to $\E\left(c_{\vecnotation{\mu},\uparrow}\right)=\E\left(c_{\vecnotation{\mu},\downarrow}\right)$. 
%Furthermore, we assume Poissonian distributed events, but take the total number of events as a constant, i.e. $N=c_{\vecnotation{\mu},\uparrow}+c_{\vecnotation{\mu},\downarrow}$ holds. Then, 
For a fixed number of counts $N=c_{\vecnotation{\mu},\uparrow}+c_{\vecnotation{\mu},\downarrow}$, the correlation becomes
\begin{equation}
\tilde T_{\vecnotation{\mu}} = \frac{1}{N}\left(c_{\vecnotation{\mu},\uparrow}-c_{\vecnotation{\mu},\downarrow}\right).
\label{eq:CorrelationSkellamDistribution}
\end{equation}
%The probability distribution of Eq.~(\ref{eq:CorrelationSkellamDistribution}) is described by the Skellam distribution~\cite{Skellam}, 
The multinomial statistics in the events $\{c_{\vecnotation{\mu},\uparrow}, c_{\vecnotation{\mu},\downarrow}\}$ tends to a normal distribution for increasing sample sizes.  
For $\E\left(c_{\vecnotation{\mu},\uparrow}\right)=\E\left(c_{\vecnotation{\mu},\downarrow}\right)=N/2$, the correlation value $T_{\vecnotation{\mu}}$ is approximately distributed according to a normal distribution with mean $0$ and standard deviation $1/\sqrt{N}$. 
Consequently, the probability density $g\left(\tilde T_{\vecnotation{\mu}}\right)$ can be expressed as
\begin{equation}
g\left(\tilde T_{\vecnotation{\mu}}\right) = \sqrt{\frac{N}{2\pi}} \exp\left(-\frac{\tilde T_{\vecnotation{\mu}}^2 N}{2}\right).
\end{equation}

Instead of a fixed number of measurements (multinomial distributed events), often the measurement time is fixed with a variable number of events.
%as in photonic experiments (Poissonian distributed events) a fixed number of measurements itself is given, we have to consider multinomially distributed counts. 
%If instead of a fixed measurement time (Poissonian distributed events) a fixed number of measurements itself is given, we have to consider multinomially distributed counts. 
%In order to illustrate the situation by means of Eq.~(\ref{eq:definition_correlation}), we use the binomial case with the probabilites $p_{\vecnotation{\mu},\uparrow}=p_{\vecnotation{\mu},\downarrow}=\frac{1}{2}$. 
Also for Poissonian distributed events and increasing sample sizes $N$, the correlation $T_{\vecnotation{\mu}}$ tends to the same normal distribution~\cite{Skellam}.
% With the help of the \textit{de Moivre-Laplace theorem}~\cite{MoivreLaplace}, one can directly show that this binomial distribution tends to the normal distribution for increasing sample sizes $N$.

In the overcomplete Pauli scheme, the expectation value of, e.g., the correlation $\sigma_{0,1}$ can be determined by means of the measurements in settings $\sigma_{1,1}$, $\sigma_{2,1}$, and $\sigma_{3,1}$. 
For those non-full correlations, more events can be taken into account, reducing its variance.
If $\sigma_\vecnotation{\mu}$ contains $j\left(\vecnotation{\mu}\right)$ factors $\sigma_0$, the distribution $g\left(\tilde T_{\vecnotation{\mu}}\right)$ is modified,
% \begin{equation}
% g\left(T_{\vecnotation{\mu}}\right) = \sqrt{\frac{N}{2\pi}} \exp\left(-\frac{T_{\vecnotation{\mu}}^2 N}{2}\right).
% \end{equation}
\begin{equation}
%\tilde T_{\vecnotation{\mu}}\sim{\cal N}\left(0,\sqrt{\frac{1}{3^{j\left(\vecnotation{\mu}\right)} N}}\right)\nonumber\\
%\Longleftrightarrow\;\;
g\left(\tilde T_{\vecnotation{\mu}}\right)=\sqrt{\frac{3^{j\left(\vecnotation{\mu}\right)} N}{2\pi}}\exp\left(-\frac{3^{j\left(\vecnotation{\mu}\right)} N \tilde T_{\vecnotation{\mu}}^2}{2}\right).
\label{eq:Correlationsprobabilitydensity}
\end{equation}
Often, the short-hand notation $j$ is used instead of $j\left(\vecnotation{\mu}\right)$.
\subsection{Limit of approximation}
Obviously, the correlation values are bounded between $-1$ and $1$, i.e., $\tilde T_{\vecnotation{\mu}}\in\left[-1,1\right]$, while the normal distribution has support everywhere in $\mathbb{R}$.
The used approximation is justified for sufficiently large $N$, i.e., for cases where enough counts are recorded such that it is fairly unlikely to obtain $|T_{\vecnotation{\mu}}|>1$ in the Gaussian approximation. 
Already for only $N=10$, the probability of $\tilde T_{\vecnotation{\mu}}\not\in\left[-1,1\right]$ is fairly low for the Gaussian approximation,
\begin{equation}
1-\int_{-1}^{1}g(\tilde T_{\vecnotation{\mu}})\rd \tilde T_{\vecnotation{\mu}}=1-\operatorname{erf}(N/2)\approx 0.0016,
\end{equation}
where $\operatorname{erf}(x)$ denotes the error function.
Thus, one can approximate the correlation with a normal distribution, where the variance ($\V\left[\tilde T_{\vecnotation{\mu}}\right]=\E\left[\tilde T_{\vecnotation{\mu}}^2\right]-\E\left[\tilde T_{\vecnotation{\mu}}\right]^2=\E\left[\tilde T_{\vecnotation{\mu}}^2\right]$ for $\E\left[\tilde T_{\vecnotation{\mu}}\right]=0$) is given by
%$\int_{1}^{\infty}g(T_{\vecnotation{\mu}})\rd T_{\vecnotation{\mu}}\approx0$ holds already for some tens of counts. 
%Thus, the assumption of 
%\begin{equation}%\begin{eqnarray}
\begin{align}
\E\left[\tilde T_{\vecnotation{\mu}}^2\right] = \int_{-1}^{1} \rd \tilde T_{\vecnotation{\mu}} g(\tilde T_{\vecnotation{\mu}}) \tilde T_{\vecnotation{\mu}}^2 %\nonumber \\
\approx \int_{-\infty}^{\infty} \rd \tilde T_{\vecnotation{\mu}} g(\tilde T_{\vecnotation{\mu}}) \tilde T_{\vecnotation{\mu}}^2\nonumber \\
= \sqrt{\frac{N}{2\pi}} \int_{-\infty}^{\infty} \rd \tilde T_{\vecnotation{\mu}} \exp\left(-\frac{N \tilde T_{\vecnotation{\mu}}^2}{2}\right) \tilde T_{\vecnotation{\mu}}^2 %\nonumber \\
= \frac{1}{N}.
% \label{eq:statisticsFullCorr}
%\end{equation}%\end{eqnarray}
\end{align}

% Show that
% \begin{itemize}
%  \item Poisson and multinomial give same results
%  \item $T\sim \CN(0,\dots\sqrt{1/N})$
%  \item limitations of low $N$ can be neglected
% \end{itemize}
\section{SM~2: Proof of semicircular distribution (Moment Method)}
This proof of the semicircular distribution is based on the moments of the correlations. 
The moments of the density matrices obtained by means of the used overcomplete Pauli scheme for quantum state tomography are compared to the moments of a semicircular distribution in order to show the equality of the distributions.
\subsection{Moments of semicircular distribution}
First, let us calculate the moments of a function describing a semicircle.
The function
\begin{equation}
f_{c,R}(x)=%	\underbrace{
\frac{2}{\pi R^2}%}_{Z}
\sqrt{R^2-\left(x-c\right)^2}
\end{equation}
describes a normalized semicircle centered around $c$ with Radius $R$. 
Without loss of generality, we focus on the central moments. 
Instead, the function $f_{c,R}(x)$ itself can be centered by setting $c=0$.
By this, moments and central moments are becoming equal.
Due to the symmetry, all odd moments vanish, i.e., $m^{\rm sc}_{2k+1}=\int_{-\infty}^{\infty}f_{0,R}\left(x\right)x^{2k+1}\rd x=0$. 
For the even moments, one finds
\begin{align}
m^{\rm sc}_{2}&=\int_{-\infty}^{\infty}f_{0,R}\left(x\right)x^{2}\rd x=\left(\frac{R}{2}\right)^2,\nonumber\\
m^{\rm sc}_{4}&=\int_{-\infty}^{\infty}f_{0,R}\left(x\right)x^{4}\rd x=2\left(\frac{R}{2}\right)^4,\nonumber\\
m^{\rm sc}_{6}&=\int_{-\infty}^{\infty}f_{0,R}\left(x\right)x^{6}\rd x=5\left(\frac{R}{2}\right)^6,\nonumber\\
m^{\rm sc}_{8}&=\int_{-\infty}^{\infty}f_{0,R}\left(x\right)x^{8}\rd x=14\left(\frac{R}{2}\right)^8,
\end{align}
where the coefficient can be found recursively and equals the \textit{Catalan numbers} ${\cal C}_{j}$.
Thus, the moments are
\begin{subequations}
\begin{alignat}{3}
%\begin{align}
m^{\rm sc}_{2k+1}=\int_{-\infty}^{\infty}f_{0,R}\left(x\right)x^{2k+1}\rd x=0,\label{eq:momentsSemicircleOdd}\\
m^{\rm sc}_{2k}=\int_{-\infty}^{\infty}f_{0,R}\left(x\right)x^{2k}\rd x=\CC_{k}\left(\frac{R}{2}\right)^{2k}=\CC_{k}\left(m^{\rm sc}_{2}\right)^k.
\label{eq:momentsSemicircleEven}
\end{alignat}
\label{eq:momentsSemicircle}
\end{subequations}
To show that the spectral probability distribution of tomographically obtained completely mixed states equals a semicircle function, one has to recover Eqs.~(\ref{eq:momentsSemicircle}) for the distribution of the eigenvalues of white noise.
\subsubsection{Catalan numbers}
The Catalan numbers are given by, e.g., a recursively defined sequence also appearing in various counting problems. 
The zeroth Catalan number is $\CC_{0}=1$.
The subsequent numbers are defined by
\begin{equation}
\CC_{j+1}=\CC_{j} \frac{2\left(2 j + 1\right)}{j+2}.
\label{eq:CatalanDefinition1}
\end{equation}
Consequently, the first elements of the (zerobased) sequence ${\cal C}_{j}$ are $1, 1, 2, 5, 14, 42, \dots$, %for $j=0,1,2,\dots$. 
counting, e.g., the possibilities of $2j$ persons shaking hands at the same time under the constraint that the hands of two pairs do not cross. 
%A famous problem featuring the $j$-th Catalan number is the question of the number of possibilities of $2j$ persons shaking hands at the same time under the constraint that the hands of two pairs do not cross. 

\subsection{Moments of density matrices}
The $k$-th moment of the eigenvalue distribution of a density matrix $\tilde \varrho$ can be found by means of 
\begin{align}%\begin{equation}%\begin{align}
m^{\rm ev}_{k}=\E\left[\frac{1}{2^n}\Tr\left(\tilde \varrho^{k}\right)\right]=\E\left[\frac{1}{2^n}\Tr\left(\left(U^{\dagger}\tilde \varrho U\right)^{k}\right)\right]\nonumber\\
=\E\left[\frac{1}{2^n}\Tr\left(D^{k}\right)\right]=\frac{1}{2^n}\E\left[\sum_{i=1}^{2^n}\lambda_i^{k}\right]=\frac{1}{2^n}\sum_{i=1}^{2^n}\E\left[\lambda_i^{k}\right],
\label{eq:eigenvaluesMoments}
\end{align}%\end{equation}%\end{align}
where $U^{\dagger}\tilde \varrho U=D$ corresponds to the eigendecomposition of $\tilde \varrho$ with diagonal $D$ ($D_{i,j}=\delta_{i,j}\lambda_i$). % consisting of the eigenvalues $\lambda_i$, i.e., $D_{i,i}=\lambda_i$ and $D_{i,j}=0$ for $i\neq j$. 
To show that the spectrum of random density matrices, namely the distribution of their eigenvalues, is semicircular, one can now prove the equality of Eq.~(\ref{eq:eigenvaluesMoments}) with the moments of a semicircle given in Eqs.~(\ref{eq:momentsSemicircle}).
% For simplicity, we assume that all correlations $T_{\vecnotation{\mu}}$ are randomly distributed according to a normal distribution with mean $\mu=0$ and standard deviation $\sigma=\sqrt{1/N}$, namely that
% \begin{align}
% &T_{\vecnotation{\mu}}\sim{\cal N}\left(0,\sqrt{\frac{1}{N}}\right)\\
% \Longleftrightarrow\;\;&g\left(T_{\vecnotation{\mu}}\right)=\sqrt{\frac{N}{2\pi}}\exp\left(-\frac{N T_{\vecnotation{\mu}}^2}{2}\right),
% \label{eq:Correlationsprobabilitydensity}
% \end{align}
% where $g\left(T_{\vecnotation{\mu}}\right)$ denotes the probability density function of $T_{\vecnotation{\mu}}$.
% As mentioned in the main text, we assume that all correlations $T_{\vecnotation{\mu}}$ are randomly distributed according to a normal distribution with mean $0$ and standard deviation $\sqrt{1/(3^j N)}$ with the number $j$ of local measurements of $\sigma_0$, namely that
% \begin{align}
% T_{\vecnotation{\mu}}\sim{\cal N}\left(0,\sqrt{\frac{1}{3^j N}}\right)\nonumber\\
% \Longleftrightarrow\;\;g\left(T_{\vecnotation{\mu}}\right)=\sqrt{\frac{3^j N}{2\pi}}\exp\left(-\frac{3^j N T_{\vecnotation{\mu}}^2}{2}\right),
% \label{eq:Correlationsprobabilitydensity}
% \end{align}
% where $g\left(T_{\vecnotation{\mu}}\right)$ denotes the probability density function of $T_{\vecnotation{\mu}}$. 
\subsubsection{First moment}
In order to calculate the central moments of the eigenvalue distribution, we need to consider
\begin{equation}
\tilde{m}^{\rm ev}_1=\E\left[\frac{1}{2^n}\Tr\left(\tilde \varrho\right)\right]=1.
\end{equation}
Instead of centralizing the moments, from now on we modify the density matrix in order to shift its spectral distribution.
By setting the constant value $T_{0,0,\dots,0}=0$, the distribution shifts its center to $0$. 
The first moment of this modified density matrix $\tilde \varrho^{\prime}=\tilde \varrho-\sigma_{0,0,\dots,0}/2^n$ is
\begin{align}
m^{\rm ev}_1=\E\left[\frac{1}{2^n}\Tr\left(\tilde \varrho-\sigma_{0,0,\dots,0}/2^n\right)\right]\nonumber\\
=\frac{1}{2^n}\E\left[\Tr\left(\tilde \varrho\right)-\Tr\left(\sigma_{0,0,\dots,0}/2^n\right)\right]=0.
\end{align}
Thus, one can calculate the moments of $\tilde \varrho^\prime$, which equal the central moments of $\tilde \varrho$.
The proof is based on the modified matrix $\tilde \varrho^\prime$, which we will from now on just call $\tilde \varrho$.
The spectral probability distributions of the original and the modified matrices only differ in the center, not in their shape.
\subsubsection{Second moment}
%Let us now calculate the expectation value of the mean value of the squared eigenvalues $m^{\rm ev}_2=1/2^n \E\left[\sum_i\lambda_i^2\right]$. 
By using Eq.~(\ref{eq:eigenvaluesMoments}) and 
%\begin{equation}
$\tilde \varrho=\frac{1}{2^n}\sum_{\vecnotation{\mu}}\tilde T_{\vecnotation{\mu}}\sigma_{\vecnotation{\mu}}$,
%\label{eq:densitymatrixparametrization}
%\end{equation}
%we obtain for the second moment $m^{\rm ev}_2$
one can calculate the second moment,
\begin{align}
m^{\rm ev}_2=\frac{1}{2^n}\sum_{i=1}^{2^n}\E\left[\lambda_i^{2}\right]=\frac{1}{2^n}\E\left[\Tr\left(\tilde \varrho^{2}\right)\right]\nonumber\\ %_{(\ref{eq:eigenvaluesMoments})}
%=\frac{1}{2^{3n}}\E\left[\Tr\left(\sum_{\vecnotation{\mu},\vecnotation{\nu}} \tilde T_{\vecnotation{\mu}}\tilde T_{\vecnotation{\nu}} \sigma_{\vecnotation{\mu}}\sigma_{\vecnotation{\nu}} \right)\right]\nonumber\\  %=_{(\ref{eq:densitymatrixparametrization})}
=\frac{1}{2^{3n}}\sum_{\vecnotation{\mu},\vecnotation{\nu}} \E\left[\tilde T_{\vecnotation{\mu}}\tilde T_{\vecnotation{\nu}}\right] \Tr\left(\sigma_{\vecnotation{\mu}}\sigma_{\vecnotation{\nu}} \right)%\nonumber\\
=\frac{2^n}{2^{3n}}\sum_{\vecnotation{\mu},\vecnotation{\nu}}  \E\left[\tilde T_{\vecnotation{\mu}}\tilde T_{\vecnotation{\nu}}\right] \delta_{\vecnotation{\mu},\vecnotation{\nu}}\label{eq:paulidirac}.
\end{align}
%where we make use of the fact that the trace for all products of tensor products of Pauli matrices vanish unless they are the same.
%where $\Tr\left(\sigma_{\vecnotation{\mu}}\sigma_{\vecnotation{\nu}} \right)=2^n\delta_{\vecnotation{\mu},\vecnotation{\nu}}$
%If for a given $\vecnotation{\mu}=(\mu_1,\dots,\mu_n)$ $j$ of the indices $\mu_k$ are $0$, $3^{n-j}\times\begin{pmatrix}n\\j\end{pmatrix}$ different correlations exist that can be used to calculate $\sigma_\vecnotation{\mu}$. %that depend on $j$ local measurements of $\sigma_0$ of $\vecnotation{\mu}$, one obtains for the second moment 
Since one can infer non-full correlations with larger statistics, see Eq.~\ref{eq:Correlationsprobabilitydensity}, one has to take this into account,
%Thus, 
%By taking into account that for a given $j$, $3^{n-j}\times\begin{pmatrix}n\\j\end{pmatrix}$ different correlations exist that depend on $j$ local measurements of $\sigma_0$ of $\vecnotation{\mu}$, one obtains for the second moment 
%For a given $j$, $3^{n-j}\times\begin{pmatrix}n\\j\end{pmatrix}$ different summands exist.
%Consequently,
\begin{align}
% \frac{1}{2^n}\sum_{i=1}^{2^n}\E\left[\lambda_i^{2}\right]
% m^{\rm ev}_2=\frac{2^n}{2^{3n}}\sum_{\vecnotation{\mu}}  \E\left[\tilde T_{\vecnotation{\mu}}^2\right] =_{(I)} \frac{4^n-1}{4^n N} \approx_{(II)} \frac{1}{N}.
m^{\rm ev}_2=\frac{2^n}{2^{3n}}\sum_{\vecnotation{\mu}}  \E\left[\tilde T_{\vecnotation{\mu}}^2\right] \nonumber \\
= \frac{1}{4^n N}\sum_{j=0}^{n-1}\begin{pmatrix}n\\j\end{pmatrix}\frac{3^{n-j}}{3^j}=\frac{10^n-1}{12^n}\frac{1}{N}\approx\frac{5^n}{6^n}\frac{1}{N}.
\label{eq:secondmoment}
\end{align}
Here, one uses that the sum runs over $\vecnotation{\mu}$, where $\E\left[\tilde T_{\mu}^2\right]=1/(3^j N)$ depends on the number $j$ of local measurements of $\sigma_0$ of $\vecnotation{\mu}$. 
%For a given $j$, $3^{n-j}\times\begin{pmatrix}n\\j\end{pmatrix}$ different summands exist.
%Proper counting leads to the given expression.
If instead of the used overcomplete Pauli scheme a tomography scheme with $\E(\tilde T_{\mu}^2)=1/N$ for all $\mu$ is used, (approximately) all matrix elements have the same variance.
In this case, the second moment is found to be $m^{\rm ev}_2=1/N$.
% In \textit{(I)} we used that the sum over $\vecnotation{\mu}$ contains $4^n-1$ non-trivial entries since the correlation $\tilde T_{0,0,\dots,0}=1$ is not considered as a random variable. 
% In the limit $n\rightarrow\infty$, the approximation \textit{(II)} can be utilized.
\subsubsection{Third and higher odd moments}
All odd (centralized) moments of the eigenvalue distribution vanish, as we will argue at the example of the third moment.
The third moment 
\begin{align}
m^{\rm ev}_3=\frac{1}{2^n}\sum_{i=1}^{2^n}\E\left[\lambda_i^{3}\right]=\frac{1}{2^n}\E\left[\Tr\left(\tilde \varrho^{3}\right)\right]\nonumber\\
=\frac{1}{2^{4n}}\E\left[\Tr\left(\sum_{\vecnotation{\mu},\vecnotation{\nu},\vecnotation{\gamma}} \tilde T_{\vecnotation{\mu}}\tilde T_{\vecnotation{\nu}}\tilde T_{\vecnotation{\gamma}} \sigma_{\vecnotation{\mu}}\sigma_{\vecnotation{\nu}}\sigma_{\vecnotation{\gamma}} \right)\right]\nonumber\\
=\frac{1}{2^{4n}}\sum_{\vecnotation{\mu},\vecnotation{\nu},\vecnotation{\gamma}} \E\left[\tilde T_{\vecnotation{\mu}}\tilde T_{\vecnotation{\nu}}\tilde T_{\vecnotation{\gamma}}\right] \Tr\left(\sigma_{\vecnotation{\mu}}\sigma_{\vecnotation{\nu}}\sigma_{\vecnotation{\gamma}} \right)
\end{align}
vanishes due to the expression $\E\left[\tilde T_{\vecnotation{\mu}}\tilde T_{\vecnotation{\nu}}\tilde T_{\vecnotation{\gamma}}\right]$.
For all choices of indices $\{\vecnotation{\mu},\vecnotation{\nu},\vecnotation{\gamma}\}$, the expecation value will be taken of at least one correlation value in an odd power, i.e., the summation contains $\E\left[\tilde T_{\vecnotation{\mu}}\tilde T_{\vecnotation{\nu}}\tilde T_{\vecnotation{\gamma}}\right]=\E\left[\tilde T_{\vecnotation{\mu}}\right]\E\left[\tilde T_{\vecnotation{\nu}}\right]\E\left[\tilde T_{\vecnotation{\gamma}}\right]$ (for mutually distinct indices $\{\vecnotation{\mu},\vecnotation{\nu},\vecnotation{\gamma}\}$), $\E\left[\tilde T_{\vecnotation{\mu}}^2\right]\E\left[\tilde T_{\vecnotation{\gamma}}\right]$ (for $\vecnotation{\mu}=\vecnotation{\nu}\neq\vecnotation{\gamma}$), and $\E\left[\tilde T_{\vecnotation{\mu}}^3\right]$ ($\vecnotation{\mu}=\vecnotation{\nu}=\vecnotation{\gamma}$).
Because of the even parity of the distribution function of the random variable [Eq.~(\ref{eq:Correlationsprobabilitydensity})], $\E\left[\tilde T_{\vecnotation{\mu}}\right]=\E\left[\tilde T_{\vecnotation{\mu}}^3\right]=0$ holds. %, i.e., the correlation and the odd parity of the correlation value appearing in odd power, the expectation value will vanish.
Thus, all odd moments vanish,  %of the (modified) density matrix, i.e., all odd centralized moments of the unmodified density matrix, have to vanish.
\begin{equation}
m^{\rm ev}_{2k+1}=0
\end{equation}
% holds $\forall k$.
\subsubsection{Fourth moment}
Extending the procedure of Eq.~(\ref{eq:secondmoment}) to the fourth moment $m^{\rm ev}_4$, one obtains
% Before we give a general expression for the moments of the spectral probability distribution of the completely mixed state, let us examine the fourth moment.
% \begin{align}
% \frac{1}{2^n}\sum_{i=1}^{2^n}\E\left[\lambda_i^{4}\right]&=&\frac{1}{2^n}&\E\left[\Tr\left(\tilde \varrho^{4}\right)\right]\nonumber\\
% &=&\frac{1}{2^{5n}}&\E\left[\Tr \sum_{\vecnotation{\mu},\vecnotation{\nu},\vecnotation{\gamma},\vecnotation{\lambda}} \tilde T_{\vecnotation{\mu}}\tilde T_{\vecnotation{\nu}}\tilde T_{\vecnotation{\gamma}}\tilde T_{\vecnotation{\lambda}} \sigma_{\vecnotation{\mu}}\sigma_{\vecnotation{\nu}}\sigma_{\vecnotation{\gamma}}\sigma_{\vecnotation{\lambda}} \right]\nonumber\\
% &=&\frac{1}{2^{5n}}&\sum_{\vecnotation{\mu},\vecnotation{\nu},\vecnotation{\gamma},\vecnotation{\lambda}} \E\left[\tilde T_{\vecnotation{\mu}}\tilde T_{\vecnotation{\nu}}\tilde T_{\vecnotation{\gamma}}\tilde T_{\vecnotation{\lambda}}\right] \cdot \nonumber \\ &&&\hspace*{1.5cm}\Tr\left(\sigma_{\vecnotation{\mu}}\sigma_{\vecnotation{\nu}}\sigma_{\vecnotation{\gamma}}\sigma_{\vecnotation{\lambda}} \right). \label{eq:fourthmoment}
% \end{align}
\begin{align}
%\frac{1}{2^n}\sum_{i=1}^{2^n}\E\left[\lambda_i^{4}\right]=\frac{1}{2^n}\E\left[\Tr\left(\tilde \varrho^{4}\right)\right]\nonumber\\
m^{\rm ev}_4=\frac{1}{2^{5n}}\E\left[\Tr \sum_{\vecnotation{\mu},\vecnotation{\nu},\vecnotation{\gamma},\vecnotation{\lambda}} \tilde T_{\vecnotation{\mu}}\tilde T_{\vecnotation{\nu}}\tilde T_{\vecnotation{\gamma}}\tilde T_{\vecnotation{\lambda}} \sigma_{\vecnotation{\mu}}\sigma_{\vecnotation{\nu}}\sigma_{\vecnotation{\gamma}}\sigma_{\vecnotation{\lambda}} \right]\nonumber\\
=\frac{1}{2^{5n}}\sum_{\vecnotation{\mu},\vecnotation{\nu},\vecnotation{\gamma},\vecnotation{\lambda}} \E\left[\tilde T_{\vecnotation{\mu}}\tilde T_{\vecnotation{\nu}}\tilde T_{\vecnotation{\gamma}}\tilde T_{\vecnotation{\lambda}}\right] %\cdot \nonumber \\
\Tr\left(\sigma_{\vecnotation{\mu}}\sigma_{\vecnotation{\nu}}\sigma_{\vecnotation{\gamma}}\sigma_{\vecnotation{\lambda}} \right). \label{eq:fourthmoment}
\end{align}
By the parity argument, only those terms contribute that contain indices in even power.
One is now left to count the number of contributing summands in Eq.~(\ref{eq:fourthmoment}). 
For illustration, consider the case of $n=2$ qubits and two different factors in the trace, each appearing twice. 
%For example, still restricting ourselves to $n=2$, we acknowledge that $\sigma_{1,1}$ and $\sigma_{2,2}$ commute, i.e., $\left[\sigma_{1,1},\sigma_{2,2}\right]=0$. 
%Thus, 
For commuting factors, e.g., $\sigma_{1,1}$ and $\sigma_{2,2}$
\begin{equation}
\Tr\left(\sigma_{1,1}\sigma_{1,1}\sigma_{2,2}\sigma_{2,2} \right) =%\nonumber\\
\Tr\left(\sigma_{1,1}\sigma_{2,2}\sigma_{1,1}\sigma_{2,2} \right) =%\nonumber \\
\dots = %\nonumber\\
%\Tr \sigma_{0,0} = 
2^2
\end{equation}
holds for all $6$ permutations of $\sigma_{1,1}\sigma_{1,1}\sigma_{2,2}\sigma_{2,2}$. 
On the other hand, one notices that for anticommuting factors $\sigma_{1,1}$ and $\sigma_{1,2}$, i.e., %$\left\{\sigma_{1,1},\sigma_{1,2}\right\}=0$, 
$\sigma_{1,1}\sigma_{1,2}=-\sigma_{1,2}\sigma_{1,1}$, 
different contributions occur.
\begin{align}
\Tr\left(\sigma_{1,1}\sigma_{1,1}\sigma_{1,2}\sigma_{1,2} \right) =%&=&\nonumber\\
\Tr\left(\sigma_{1,1}\sigma_{1,2}\sigma_{1,2}\sigma_{1,1} \right) =\nonumber \\%&=&\nonumber \\
\Tr\left(\sigma_{1,2}\sigma_{1,1}\sigma_{1,1}\sigma_{1,2} \right) =%&=&\nonumber \\
\Tr\left(\sigma_{1,2}\sigma_{1,2}\sigma_{1,1}\sigma_{1,1} \right) =%&=&\nonumber \\
2^2%\Tr \sigma_{0,0}=4%&=&4
\end{align}
and
\begin{equation}
\Tr\left(\sigma_{1,1}\sigma_{1,2}\sigma_{1,1}\sigma_{1,2} \right) =%&=&\nonumber\\
\Tr\left(\sigma_{1,2}\sigma_{1,1}\sigma_{1,2}\sigma_{1,1} \right) =%&=&\nonumber \\
%-\Tr \sigma_{0,0}=-4%&=&-4
-2^2
\end{equation}
are valid for the permutations of two \textit{anti}commuting factors, each of those appearing twice.
Hence, if the matrices $\sigma_{\vecnotation{\mu}}$ and $\sigma_{\vecnotation{\nu}}$ commute, $6$ terms are contributing while anticommuting $\sigma_{\vecnotation{\mu}}$ and $\sigma_{\vecnotation{\nu}}$ lead to an effective contribution of only $4-2=2$ summands. 

We already could argue that the four indices $\vecnotation{\mu},\vecnotation{\nu},\vecnotation{\gamma},\vecnotation{\lambda}$ in Eq.~(\ref{eq:fourthmoment}) reduce to two indices, e.g. $\vecnotation{\mu}$ and $\vecnotation{\nu}$, where the sum in Eq.~(\ref{eq:fourthmoment}) is now running over those two indices and all permutations of $\sigma_{\vecnotation{\mu}}\sigma_{\vecnotation{\mu}}\sigma_{\vecnotation{\nu}}\sigma_{\vecnotation{\nu}}$. 
By a simple counting argument, one realizes that after choosing $\vecnotation{\mu}$ out of approximately $4^n$ possibilities, one has approximately $4^n/2$ possibilities for $\vecnotation{\nu}$ such that $\left[\sigma_{\vecnotation{\mu}},\sigma_{\vecnotation{\nu}}\right]=0$, while the other $4^n/2$ choices of $\vecnotation{\nu}$ lead to $\left\{\sigma_{\vecnotation{\mu}},\sigma_{\vecnotation{\nu}}\right\}=0$.
The expressions
\begin{align}
\Tr\left(\sigma_{\vecnotation{\mu}}\sigma_{\vecnotation{\mu}}\sigma_{\vecnotation{\nu}}\sigma_{\vecnotation{\nu}} \right) =%&=& \nonumber\\
\Tr\left(\sigma_{\vecnotation{\mu}}\sigma_{\vecnotation{\nu}}\sigma_{\vecnotation{\nu}}\sigma_{\vecnotation{\mu}} \right) =\nonumber \\%&=& \nonumber \\
\Tr\left(\sigma_{\vecnotation{\nu}}\sigma_{\vecnotation{\mu}}\sigma_{\vecnotation{\mu}}\sigma_{\vecnotation{\nu}} \right) =%&=& \nonumber \\
\Tr\left(\sigma_{\vecnotation{\nu}}\sigma_{\vecnotation{\nu}}\sigma_{\vecnotation{\mu}}\sigma_{\vecnotation{\mu}} \right) =%&=& \nonumber \\
2^2%\Tr \sigma_{0,0}=4%&=& 4
\end{align}
give a positive contribution independently of the commutation relation between $\sigma_{\vecnotation{\mu}}$ and $\sigma_{\vecnotation{\nu}}$.
%In contrast,
\begin{equation}
\Tr\left(\sigma_{\vecnotation{\mu}}\sigma_{\vecnotation{\nu}}\sigma_{\vecnotation{\mu}}\sigma_{\vecnotation{\nu}} \right) =%&=& \nonumber\\
\Tr\left(\sigma_{\vecnotation{\nu}}\sigma_{\vecnotation{\mu}}\sigma_{\vecnotation{\nu}}\sigma_{\vecnotation{\mu}} \right) =%&=& \nonumber \\
\pm\Tr \sigma_{0,0}=\pm2^2.%&=& \pm4.
\end{equation}
has in half of the choices of $\vecnotation{\mu}$ and $\vecnotation{\nu}$ positive, in the other half negative contribution.
Contributions of those permutations cancel for different choices of indices. %, they do not have to be considered, %these permutations do not have to be considered, 
% \onecolumngrid
\begin{widetext}
\begin{subequations}
\begin{alignat}{3}
m^{\rm ev}_4=\frac{1}{2^n}\sum_{i=1}^{2^n}\E\left[\lambda_i^{4}\right]=%\nonumber\\
\frac{1}{2^{5n}}\sum_{\vecnotation{\mu},\vecnotation{\nu},\vecnotation{\gamma},\vecnotation{\lambda}} \E\left[\tilde T_{\vecnotation{\mu}}\tilde T_{\vecnotation{\nu}}\tilde T_{\vecnotation{\gamma}}\tilde T_{\vecnotation{\lambda}}\right] \Tr\left(\sigma_{\vecnotation{\mu}}\sigma_{\vecnotation{\nu}}\sigma_{\vecnotation{\gamma}}\sigma_{\vecnotation{\lambda}} \right) = \nonumber\\
\frac{1}{2^{5n}}\frac{1}{2!}\sum_{\vecnotation{\mu}}\Bigg[\sum_{\vecnotation{\nu}:\{\vecnotation{\nu}\neq\vecnotation{\mu}\}} \E\left[\tilde T_{\vecnotation{\mu}}^2\tilde T_{\vecnotation{\nu}}^2\right] \Tr\left(\sum_{i=1}^{6}{\cal P}_{i}\left(\sigma_{\vecnotation{\mu}}\sigma_{\vecnotation{\mu}}\sigma_{\vecnotation{\nu}}\sigma_{\vecnotation{\nu}} \right)\right) %\nonumber \\
 + \E\left[\tilde T_{\vecnotation{\mu}}^4\right] \Tr\left(\sigma_{\vecnotation{\mu}}\sigma_{\vecnotation{\mu}}\sigma_{\vecnotation{\mu}}\sigma_{\vecnotation{\mu}} \right) \Bigg] \approx \label{eq:fourthmomentPermutations} \\
\frac{1}{2^{5n}}\frac{1}{2!}\sum_{\vecnotation{\mu}}\sum_{\vecnotation{\nu}:\{\vecnotation{\nu}\neq\vecnotation{\mu}\}} \E\left[\tilde T_{\vecnotation{\mu}}^2\tilde T_{\vecnotation{\nu}}^2\right] \Tr\left(\sum_{i=1}^{6}{\cal P}_{i}\left(\sigma_{\vecnotation{\mu}}\sigma_{\vecnotation{\mu}}\sigma_{\vecnotation{\nu}}\sigma_{\vecnotation{\nu}} \right)\right),
\end{alignat}
\end{subequations}
where 
%\begin{equation}
$\sum_{i=1}^{6}{\cal P}_{i}\left(\sigma_{\vecnotation{\mu}}\sigma_{\vecnotation{\mu}}\sigma_{\vecnotation{\nu}}\sigma_{\vecnotation{\nu}} \right)=%\nonumber\\
\sigma_{\vecnotation{\mu}}\sigma_{\vecnotation{\mu}}\sigma_{\vecnotation{\nu}}\sigma_{\vecnotation{\nu}}+
\sigma_{\vecnotation{\mu}}\sigma_{\vecnotation{\nu}}\sigma_{\vecnotation{\mu}}\sigma_{\vecnotation{\nu}}+
%\sigma_{\vecnotation{\mu}}\sigma_{\vecnotation{\nu}}\sigma_{\vecnotation{\nu}}\sigma_{\vecnotation{\mu}}+
\dots+
\sigma_{\vecnotation{\nu}}\sigma_{\vecnotation{\nu}}\sigma_{\vecnotation{\mu}}\sigma_{\vecnotation{\mu}}
$ %\end{equation}
denotes the summation over all permutations of the tensor products of Pauli matrices. 
% By means of Eq.~(\ref{eq:Correlationsprobabilitydensity}), one finds $\E\left[\tilde T_{\vecnotation{\mu}}^2\tilde T_{\vecnotation{\nu}}^2\right]=1/N^2$ and $\E\left[\tilde T_{\vecnotation{\mu}}^4\right]=3/N^2$.
By means of Eq.~(\ref{eq:Correlationsprobabilitydensity}), one finds $\E\left[\tilde T_{\vecnotation{\mu}}^2\tilde T_{\vecnotation{\nu}}^2\right]=\E\left[\tilde T_{\vecnotation{\mu}}^2\right]\E\left[\tilde T_{\vecnotation{\nu}}^2\right]=1/\left(3^{j_\vecnotation{\mu}}N\right)1/\left(3^{k_\vecnotation{\nu}}N\right)$ and $\E\left[\tilde T_{\vecnotation{\mu}}^4\right]=3/\left(3^{j_\vecnotation{\mu}} N\right)^2$, where $j_\vecnotation{\mu}$ and $k_\vecnotation{\nu}$ denote the number of $\sigma_0$ operators appearing in $\sigma_\vecnotation{\mu}$ and $\sigma_\vecnotation{\nu}$, respectively.
Because the first summand in Eq.~(\ref{eq:fourthmomentPermutations}) is occurring ${\cal O}(4^n)$ times more often, the second term can be neglected.
% The second summand in Eq.~(\ref{eq:fourthmomentPermutations}) can be neglected in comparison to the first summand occurring ${\cal O}(4^n)$ more often. 
With the aforementioned argumentation, $\Tr\left(\sum_{i=1}^{6}{\cal P}_{i}\left(\sigma_{\vecnotation{\mu}}\sigma_{\vecnotation{\mu}}\sigma_{\vecnotation{\nu}}\sigma_{\vecnotation{\nu}} \right)\right)=4\cdot2^n$ holds, leading to
\begin{equation}
m^{\rm ev}_4=\frac{1}{2^n}\sum_{i=1}^{2^n}\E\left[\lambda_i^{4}\right]=%\nonumber\\
\frac{1}{2}\frac{1}{2^{5n}}\sum_{\vecnotation{\mu}}\sum_{\vecnotation{\nu}:\{\vecnotation{\nu}\neq\vecnotation{\mu}\}} \frac{1}{N^2} 4\cdot2^n=%\nonumber\\
%\frac{1}{2}\frac{1}{2^{5n}}4^n 4^n \frac{1}{N^2} 4\cdot2^n =\nonumber\\
\frac{1}{2}\frac{1}{2^{5n}}4^n 4^n \left(\frac{5^{n}}{6^{n}N}\right)^2 4\cdot2^n =%\nonumber\\
2\cdot\left(\frac{5^{n}}{6^{n}N}\right)^2=2 \left(m^{\rm ev}_2\right)^2.
\end{equation}
Using instead a tomography scheme with $\E(\tilde T_{\vecnotation{\mu}}^2)=1/N$ $\forall \vecnotation{\mu}$ leads to $m^{\rm ev}_4=2/N^2$.
\subsubsection{Sixth moment}
Analogously, the sixth moment reads
\begin{equation}
m^{\rm ev}_6=\frac{1}{2^n}\sum_{i=1}^{2^n}\E\left[\lambda_i^{6}\right]\approx%\nonumber\\
\frac{1}{2^{7n}}\frac{1}{3!}\sum_{\vecnotation{\mu}}\sum_{\vecnotation{\nu}:\{\vecnotation{\nu}\neq\vecnotation{\mu}\}}\sum_{\vecnotation{\gamma}:\{\vecnotation{\gamma}\neq\vecnotation{\nu},\vecnotation{\gamma}\neq\vecnotation{\mu}\}}\E\left[\tilde T_{\vecnotation{\nu}}^2\tilde T_{\vecnotation{\mu}}^2\tilde T_{\vecnotation{\gamma}}^2\right]\cdot%\nonumber\\
\Tr\left(\sum_{i=1}^{90}{\cal P}_{i}\left(\sigma_{\vecnotation{\mu}}\sigma_{\vecnotation{\mu}}\sigma_{\vecnotation{\nu}}\sigma_{\vecnotation{\nu}}\sigma_{\vecnotation{\gamma}}\sigma_{\vecnotation{\gamma}} \right)\right),
\label{eq:sixthmoment}
\end{equation}
where in the approximation terms of the form $\E\left[\tilde T_{\vecnotation{\nu}}^4\tilde T_{\vecnotation{\mu}}^2\right]$ and $\E\left[\tilde T_{\vecnotation{\nu}}^6\right]$ are neglected. 
The factor $1/3!$ compensates for multiple counting of exchanging indices.
A specific summand of that expression, e.g.,
% For a given $\vecnotation{\mu}$ (out of approximately $4^n$), $\vecnotation{\nu}$ is choosen out of $4^n/2$ possibilities because of $\vecnotation{\nu}<\vecnotation{\mu}$. 
% By the restriction of $\vecnotation{\gamma}<\vecnotation{\nu}$ to avoid multiple counting, there are $4^n/4$ choices for $\vecnotation{\gamma}$. 
% Consider now, e.g., the expression 
%\begin{equation}
$\Tr\left(\sigma_{\vecnotation{\mu}}\sigma_{\vecnotation{\nu}}\sigma_{\vecnotation{\gamma}}\sigma_{\vecnotation{\mu}}\sigma_{\vecnotation{\nu}}\sigma_{\vecnotation{\gamma}} \right)$ 
%\label{eq:moments6Crossing}
%\end{equation}
% Eq.~(\ref{eq:moments6Crossing}) 
contributes positively for half of the choices of $\{\vecnotation{\mu},\vecnotation{\nu},\vecnotation{\gamma}\}$, while the other choices lead to a negative contribution.
Thus, permutations of that form %of Eq.~(\ref{eq:moments6Crossing}) 
cancel each other and can thus be neglected.
By extending this argument, all \textit{crossing} partitions do not have to be considered.
Hence, one has to count only \textit{noncrossing} partitions.
Consequently, Eq.~(\ref{eq:sixthmoment}) can be further simplified leading to %we can further simplify and obtain by properly counting
\begin{equation}
m^{\rm ev}_6=\frac{1}{2^n}\sum_{i=1}^{2^n}\E\left[\lambda_i^{6}\right]\approx%\nonumber\\
%\frac{1}{2^{7n}}\frac{1}{3!}4^n \cdot 4^n\cdot4^n \frac{1}{N^3} 2^n 30=\frac{5}{N^3}=5 \left(m^{\rm ev}_4\right)^3.
\frac{1}{2^{7n}}\frac{1}{3!}4^n \cdot 4^n\cdot4^n \left(\frac{5^{n}}{6^{n}N}\right)^3 2^n 30=%\nonumber\\
5\left(\frac{5^{n}}{6^{n}N}\right)^3=5 \left(m^{\rm ev}_2\right)^3.
\label{eq:sixthmoment2}
\end{equation}
For a scheme with $\E(\tilde T_{\mu}^2)=1/N$ $\forall \mu$, one obtains $m^{\rm ev}_6=5/N^3$.
% \twocolumngrid
\end{widetext}

\subsubsection{Crossing and noncrossing partitions}
\begin{figure}
\includegraphics[width=0.48\textwidth]{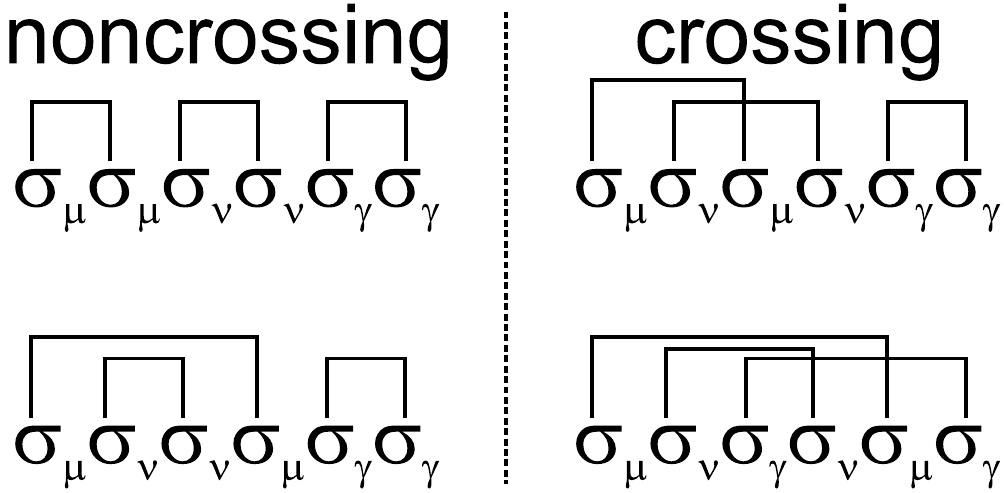}
\caption{Examples of \textit{crossing} and \textit{noncrossing} partitions. 
Four different permutations of $\sigma_{\vecnotation{\mu}}\sigma_{\vecnotation{\mu}}\sigma_{\vecnotation{\nu}}\sigma_{\vecnotation{\nu}}\sigma_{\vecnotation{\gamma}}\sigma_{\vecnotation{\gamma}}$ are depicted, where equal operators are connected by lines. 
The permutations in the left column are called \textit{noncrossing}, which corresponds to the graphical representation of noncrossing lines. 
Thus, the trace of those products is independent whether $\sigma_{\vecnotation{\mu}}$, $\sigma_{\vecnotation{\nu}}$ and $\sigma_{\vecnotation{\gamma}}$ mutually commute or anticommute.
In contrast, the trace of permutations shown in the right column depends on the (anti-)commutation relations. 
E.g., $\Tr\left(\sigma_{\vecnotation{\mu}}\sigma_{\vecnotation{\nu}}\sigma_{\vecnotation{\mu}}\sigma_{\vecnotation{\nu}}\sigma_{\vecnotation{\gamma}}\sigma_{\vecnotation{\gamma}} \right)$ is positive if $\left[\sigma_{\vecnotation{\mu}},\sigma_{\vecnotation{\nu}}\right]=0$. 
These terms are posivitely contributing to the sum in Eq.~(\ref{eq:sixthmoment}), but cancel with the negative values for $\vecnotation{\nu}$ with $\left\{\sigma_{\vecnotation{\mu}},\sigma_{\vecnotation{\nu}}\right\}=0$.
}
\label{fig:partitions}
\end{figure}
Fig.~\ref{fig:partitions} illustrates four examples of permutations occuring in the calculation of the sixth moment in Eq.~(\ref{eq:sixthmoment}).
Only noncrossing permutations, where the (anti-)commutation relations between the operators do not influence the value of the trace, have to be considered.
Thus, we are left to count the number of noncrossing partitions.
For our case of $k/2$ different operators, each of those appearing twice, the number of noncrossing partitions is given by the \textit{Catalan number} $\CC_{k/2}$, see Eq.~(\ref{eq:CatalanDefinition1}).
% \begin{equation}
% \CC_{\frac{k}{2}}=\frac{1}{\frac{k}{2}+1} 
%  \begin{pmatrix}
% k\\\frac{k}{2}
%  \end{pmatrix},
% \label{eq:CatalanDefinition2}
% \end{equation}
% which is another expression defining the \textit{Catalan numbers}, fully equivalent to Eq.~(\ref{eq:CatalanDefinition1}).\\
Hence, all odd moments vanish while for the $2k$-th moment
\begin{equation}
m^{\rm ev}_{2k}=\frac{1}{2^n}\sum_{i=1}^{2^n}\E\left[\lambda_i^{2k}\right]=\CC_k \left(\frac{5^n}{6^nN}\right)^k
\end{equation}
holds, which equals the expressions of Eqs.~(\ref{eq:momentsSemicircle}) for $R=2\left(5/6\right)^{\left(n/2\right)}/\sqrt{N}$. 
% Please note that this radius is obtained under the assumption that all random variables, namely correlations, have the same standard deviation. 
% For the concretely used tomography scheme, this relation is modified, see main text.
Eventually, we could prove that the spectral probability distribution of the completely mixed state converges to the Wigner semicircle in the limit of many qubits.
For a tomography scheme with equal variances for all correlations (and all matrix elements), one obtains $R=2\sqrt{1/N}$.

%\section{SM~3: Derivation of radius equation}
%In the main text, we made use of the relationship between the expectation value of the trace of the $2k$-th power of the density matrix with the radius, i.e. of 
%\begin{equation}
%\E\left[\frac{1}{2^n}\Tr\left(\tilde \varrho^{2 k}\right)\right]=\left(\frac{R}{2}\right)^{2 k} \CC_k
%\label{eq:moments}
%\end{equation}
%with the Catalan numbers $\CC_k$ to obtain for $k=1$ the following relationship
%\begin{equation}
%\frac{1}{4^n} \sum_{\vecnotation{\mu}} \E\left( \tilde T_{\vecnotation{\mu}}^2 \right)=\left(\frac{R}{2}\right)^{2}.
%\label{eq:momentsk1}
%\end{equation}
%Eq.~\ref{eq:moments} can easily be obtained by integrating over a semicircle. With the normalization
%\begin{equation}
%Z = \int_{-R}^{R} \sqrt{R^2-x^2} \D x = R^2 \frac{\pi}{2}
%\end{equation}
%we obtain
%\begin{align}
%\frac{1}{Z} \int_{-R}^{R} x^2 \sqrt{R^2-x^2} \D x \nonumber\\
%=\frac{1}{Z} \frac{R^4 \pi}{8} = \left(\frac{R}{2}\right)^2.
%\end{align}
%Hence, the second (centralized) moment of a semicircular distribution is proportional to the squared radius.
%Also the general Eq.~\ref{eq:moments} can be derived by integration, i.e.
%\begin{align}
%\frac{1}{Z} \int_{-R}^{R} x^{2k} \sqrt{R^2-x^2} \D x = \left(\frac{R}{2}\right)^{2k} \underbrace{\frac{4^k \operatorname{\Gamma}(k+\frac{1}{2})}{\sqrt{\pi}\operatorname{\Gamma}(k+2)}}_{\CC_k}.
%\end{align}
%For a full derivation of Eq.~\ref{eq:moments}, one has to show that $\frac{4^k \operatorname{\Gamma}(k+\frac{1}{2})}{\sqrt{\pi}\operatorname{\Gamma}(k+2)}=\CC_k$ holds for $k\in\N$, which can easily be proved by means of induction.

\section{SM~3: Proof of semicircular distribution (Matrix Theory)}\label{sec:geom_set}\label{sec:theorychapter}
We use methods of random matrix theory to show that in the limit of large density matrices, i.e. many qubits, a Wigner semi circle distribution is expected for the eigenvalues of density matrices. 
This proof, based on standard textbooks, uses the assumption that all matrix elements have equal variances.
Thus, it considers a special kind of tomography scheme. 
For specific details of the informationally overcomplete Pauli tomography method, see SM~2.
%\textcolor{red}{Some parts of this proof (SM~1) are to be revised.: \textit{Our proof does not rely on a specific tomography scheme for obtaining the density matrices.} How does this work}
%Thus, this proof is kept general. 
%The proof follows standard textbooks, but is applied to density matrices.

%Here, we prove the semicircular distribution of eigenvalues of density matrices in the limit of many qubits.
%
\subsection{Definition}
The partition function of matrix models is defined as~\cite{Marino,Mehta}
\begin{align}
Z=\frac{1}{\mathrm{vol}(U(\tilde{n}))} \int \D \tilde \varrho e^{-\frac{1}{g_s} W(\tilde \varrho)}
\end{align}
where $\tilde \varrho$ is a $\tilde{n}\times \tilde{n}$ matrix with $\tilde{n}=2^n$ with $n$ denoting the number of qubits, $\mathrm{vol}(U(\tilde{n}))$ the volume of the gauge group of the theory and $W(\tilde \varrho)$ the action.

We use the $U(\tilde{n})$ gauge freedom to set 
% \begin{align}
$\tilde \varrho\rightarrow D=U \tilde \varrho U^\dagger$ 
% \end{align}
with the diagonal matrix $D=\mathrm{diag}(\lambda_1,\dots,\lambda_{\tilde{n}})$. Using the Faddeev-Popov trick, we get for the gauge fixed path integral
\begin{align}
Z=\CN \int \prod_{i=1}^{\tilde{n}} \D \lambda_i \det \left ( \frac{\delta F (^U \tilde \varrho)}{\delta A} \right ) e^{-\frac{1}{g_s} W(\lambda)}
\end{align}
where $F=0$ is the gauge fixing condition and $U=e^A$. Calculating the Faddeev-Popov determinant with $F_{ij}(^U D)=(U D U^\dagger )_{i j}=A_{ij}=(\lambda_i -\lambda_j)+\dots$
\begin{align}
\Delta ^2(\lambda)=\det \left ( \frac{\delta F (^U \tilde \varrho)}{\delta A} \right ) =\prod_{i<j}(\lambda_i -\lambda_j)^2\; .
\end{align}
After fixing the normalization one finds for the partition function in the eigenvalue representation
\begin{align}\label{eq:eigenvalueform}
Z=\frac{1}{{\tilde{n}}! (2 \pi)^{\tilde{n}}} \int \prod_{i=1}^{\tilde{n}} \D \lambda_i \prod _{i<j}(\lambda_i -\lambda_j)^2 e^{-\frac{1}{g_s} \sum_i W(\lambda_i)}
\end{align}

\subsection{Potential for density matrices}
The action of the matrix model is a power series% in traces
\begin{align}
\frac{1}{g_s} W(\tilde \varrho)=\frac{1}{2g_s} \tr \tilde \varrho^2 +\frac{1}{g_s}\sum_{p\ge3}\frac{g_p}{p}\tr \tilde \varrho^p.
\end{align}
%The question now is: what is the right action for density matrices? 
We are now looking for the right action for density matrices. 
Since our aim is to compute the large $\tilde{n}$ limit of the theory, it suffices to answer this question as well only in this limit. For density matrices the constraints
\begin{align}
\tr \tilde \varrho &= \sum_i \lambda_i = 1\\
\tr \tilde \varrho^2 &= \sum_i \lambda_i^2\le 1
\end{align}
are fulfilled. 
% After transforming, i.e. gauging, the density matrix to its eigenvalue basis, we can reformulate these constraints for the eigenvalues $\lambda_i$
% \begin{align}
% \sum_i \lambda_i &=1\\
% \sum_i \lambda_i^2&\le 1\; .
% \end{align}
%\emph{Since for density matrices with a physical representation every eigenvalue corresponds to a probability, the eigenvalues have to fulfill $0\le \lambda_i \le 1$} --- \textit{We have to omit this statement since we want to skip the physicality constraint!}. 
Since we do not want to constrain our findings to physical density matrices, where $0\le \lambda_i \le 1$ holds, we only make use of the fact that the eigenvalues fulfill $|\lambda_i| \le 1$. 
Consequently, a generic eigenvalue behaves as $\CO (\frac{1}{\tilde{n}})$. Using that to determine the behaviour of generic traces, we find
%\begin{align}
% \tr \tilde \varrho^k &\propto \sum \frac{1}{\tilde{n}^k}\\
% &\propto \CO \left(\frac{1}{\tilde{n}^{k-1}}\right) \; .
$\tr \tilde \varrho^k \propto \sum \frac{1}{\tilde{n}^k} \propto \CO \left(\frac{1}{\tilde{n}^{k-1}}\right)$. % \; .
%\end{align}
Because non-generic density matrices have vanishing density in the whole set of matrices, we can assume the action in the limit of many qubits, i.e. of large $\tilde{n}$, to be 
% \begin{align}
$\frac{1}{g_s} W(\tilde \varrho)=\frac{1}{2g_s} \tr \tilde \varrho^2$. % \; .
% \end{align}
%which is constant in $\tilde{n}$. 
The term including $\tr \tilde \varrho$ is constant and thus leads only to a normalization factor for the partition function.

\subsection{Eigenvalue distribution}
To solve the matrix model we use a saddlepoint analysis~\cite{Marino,Mehta}. The eigenvalue description from Eq.~(\ref{eq:eigenvalueform}) can be rewritten as
\begin{align}
Z=\frac{1}{\tilde{n}!} \int \prod _{i=1}^{\tilde{n}} \frac{\D \lambda_i}{2\pi} e^{{\tilde{n}}^2 S_{\mathrm{eff}}(\lambda)}
\end{align}
with the effective action defined as
\begin{align}
S_{\mathrm{eff}}(\lambda) = -\frac{1}{t \tilde{n}} \sum_{i=1}^{\tilde{n}} W(\lambda_i)+\frac{2}{\tilde{n}^2}\sum_{i<j} \log |\lambda_i-\lambda_j|
\end{align}
where $t = \tilde{n} g_s$. 
Since the effective action is $\CO(1)$ in $\tilde{n}$ as $\tilde{n}\rightarrow \infty$, $t$ constant, we find the saddle point
\begin{align}\label{eq:discrpotential}
\frac{1}{2 t} W^\prime (\lambda_i) = \frac{1}{\tilde{n}} \sum_{j\ne i}\frac{1}{\lambda_i-\lambda_j}
\end{align}
for $i=1,\dots,\tilde{n}$.

Our aim is to calculate the eigenvalue distribution, which is defined as 
% \begin{align}
$\density (\lambda )= \frac{1}{\tilde{n}}\sum_{i=1}^{\tilde{n}} \delta (\lambda-\lambda_i)$. % \; .
% \end{align}
In the large $\tilde{n}$ limit we have to find the continuum limit of this, which can be found by using 
\begin{align}
\frac{1}{\tilde{n}} \sum_{i=1}^{\tilde{n}} f(\lambda_i)\rightarrow \int f(\lambda) \density (\lambda) \D \lambda 
\end{align}
with the normalisation constraint 
% \begin{align}
$\int \density (\lambda) \D \lambda = 1$. % \; .
% \end{align}
Using the above continuum limit procedure, we find as the continuum limit of Eq.~(\ref{eq:discrpotential})
\begin{align}
\frac{1}{2 t} W^\prime (\lambda)=\mathrm P \int_\CC \frac{\density(\lambda^\prime) \D \lambda^\prime}{\lambda-\lambda^\prime} \; .
\end{align}
The resolvent is defined as
\begin{align}
\omega (p) = \frac{1}{\tilde{n}} \left< \tr \frac{1}{p-\tilde \varrho} \right> \; .
\end{align}
In the large $\tilde{n}$ limit this can be written as
\begin{align}
\omega _0 (p) = \int \D \lambda \frac{\density (\lambda)}{p-\lambda} \; .
\end{align}
Asymptotically, the resolvent behaves as
\begin{align}\label{eq:asymptotic}
\omega _0 (p)\sim \frac{1}{p}, \quad p\rightarrow \infty \; .
\end{align}
As a function of $p$, $\omega _0 (p)$ has a discontinuity along a cut $\CC$ where the eigenvalues are located. 
The discontinuity can be calculated by means of 
\begin{align}
\omega_0(p+i \epsilon)&=\int_\mathbb{R} \D \lambda \frac{\density(\lambda)}{p+i \epsilon-\lambda}\\
&=\int_{\mathbb{R}- i \epsilon} \D \lambda \frac{\density(\lambda)}{p-\lambda}\\
&= \mathrm{P}\int \D \lambda \frac{\density(\lambda)}{p-\lambda} + \int _{C_\epsilon} \D \lambda \frac{\density(\lambda)}{p-\lambda}\\
&= \mathrm{P}\int \D \lambda \frac{\density(\lambda)}{p-\lambda}-\pi i \density(p)
\end{align}
where the contour $C_\epsilon$ goes counterclockwise around the point $\lambda=p$ in the lower half plane and $\mathrm{P}$ denotes the principal value. 
Analogously, we get
\begin{align}
\omega_0(p-i \epsilon)=\mathrm{P}\int \D \lambda \frac{\density(\lambda)}{p-\lambda} +\pi i \density(p) \; .
\end{align}
Subtracting these two expressions, we can compute the eigenvalue density from the resolvent
\begin{align}\label{eq:eigenvaluedensity}
\density(\lambda)=-\frac{1}{2\pi i}(\omega_0(\lambda+i \epsilon) - \omega_0(\lambda-i \epsilon)) \; .
\end{align}

\subsection{Calculating the resolvent}
In order to calculate the resolvent, we use again the discontinuity by adding the two terms
\begin{align}
% \omega_0(\lambda+i \epsilon) + \omega_0(\lambda-i \epsilon))&=2\mathrm P \int \D \lambda^\prime \frac{\density(\lambda^\prime)}{\lambda-\lambda^\prime}\\
% &=\frac{1}{t}W^\prime(\lambda)
\omega_0(\lambda+i \epsilon) + \omega_0(\lambda-i \epsilon))=2\mathrm P \int \D \lambda^\prime \frac{\density(\lambda^\prime)}{\lambda-\lambda^\prime}=\frac{1}{t}W^\prime(\lambda)
\end{align}
This equation can be solved by~\cite{Migdal,Marino}
\begin{align}\label{eq:omega}
\omega_0(p)=\frac{1}{2t}\oint _\CC \frac{\D z}{2\pi i} \frac{W^\prime(z)}{p-z}\left ( \frac{(p-a)(p-b)}{(z-a)(z-b)}\right )^\frac{1}{2}
\end{align}
where $a$ and $b$ are the boundaries of $\CC$ such that $b\le \lambda \le a$ for $\lambda \in \CC$.

The boundaries $a$ and $b$ can be calculated by use of the asympotic behaviour of Eq.~(\ref{eq:omega}) and Eq.~(\ref{eq:asymptotic})
\begin{align}
\oint_\CC \frac{\D z}{2\pi i}\frac{W^\prime(z)}{\sqrt{(z-a)(z-b)}} &= 0 \label{eq:bdry1}\\
\oint_\CC \frac{\D z}{2\pi i}\frac{z W^\prime(z)}{\sqrt{(z-a)(z-b)}} &= 2t \label{eq:bdry2}\; .
\end{align}
Eq.~(\ref{eq:omega}) can be simplified for polynomial potentials by a deformation of the contour. In this process we pick up poles at $\infty$ and at $p$ and thus get
\begin{align}
\omega_0(p)&=\frac{1}{2t}W^\prime(p)-\frac{1}{2t}\sqrt{(p-a)(p-b)}\tilde \varrho(p)\; ,\\
M(p)&=\oint_0 \frac{\D z}{2\pi i} \frac{W^\prime (1/z)}{1-pz}\frac{1}{\sqrt{(1-az)(a-bz)}} \; .
\end{align}

\subsection{Eigenvalue distribution for our example}
The action for density matrices in the large $\tilde{n}$ limit was found to be $W(\tilde \varrho)=\frac{1}{2} \Tr \tilde \varrho^2$. 
In terms of the eigenvalues, the action can be expressed as $W(\lambda)=\frac{1}{2} \lambda^2$, leading to $W^\prime (\lambda)=\lambda$.

%For the eigenvalue action this means $W(\lambda)=\frac{1}{2} \lambda^2$ and thus $W^\prime (\lambda)=\lambda$.\\
Plugging this into Eq.~(\ref{eq:bdry1}) and deforming the contour to infinity, we find
\begin{align}
0&=\oint _\CC \frac{\D z}{2\pi i}\frac{z}{\sqrt{(z-a)(z-b)}}\label{eq:pathintegral1} \\
&=\oint _0 \frac{\D y}{2\pi i y^2}\frac{1/y}{\sqrt{1/y^2 (1-ay)(1-by)}}\label{eq:pathintegral2}\\
&=\frac{a+b}{2}\label{eq:pathintegral3}
\end{align}
where we used $y=\frac{1}{z}$ in line Eq.~(\ref{eq:pathintegral2}). Applying the same procedure, we find in Eq.~(\ref{eq:bdry2})  
% \begin{align}
$a=2\sqrt{t}$. % \; .
% \end{align}
With this we can calculate $\tilde \varrho(p)$ and $\omega_0 (p)$,
\begin{align}
\tilde \varrho(p)&=1\\
\omega_0(p)&=\frac{1}{2t}(p-\sqrt{p^2-4t}) \; .
\end{align}
Plugging this into Eq.~(\ref{eq:eigenvaluedensity}), i.e. the equation for the eigenvalue density, one finds
\begin{align}
\density(\lambda)&=\frac{1}{2\pi t} \sqrt{4 t -\lambda ^2}
\end{align}
which is the Wigner semicircle law with radius $2 \sqrt{t}$. 
%We use this asymptotic description of $\tilde{n}=2^n\rightarrow\infty$ also for cases with only a few qubits $n$, where we calculate the radius $R=2 \sqrt{t}$ by means of the second moment of the distribution, see SM~3.

\section{SM~4: Spectral probability distribution of one qubit density matrices}
%This section details how to derive the spectral probability distribution in the single qubit case as stated in Eq.~(\ref{eq:analytical_solution_1qubit_wn}). 
The spectral probability distribution for the single qubit case as stated in Eq.~(\ref{eq:analytical_solution_1qubit_wn}) can be derived analytically.
The eigenvalues are given by 
%\begin{equation}
$\lambda_{1,2} = \frac{1}{2} \left(1 \pm \sqrt{\tilde T_1^2+\tilde T_2^2+\tilde T_3^2}\right)$.
%\end{equation}
Using the assumption of Gaussian distributed correlations, one obtains for the density $g$ of eigenvalues $\lambda$ 
% \onecolumngrid
\begin{widetext}
\begin{equation}
g(\lambda) = \int_{-\infty}^{\infty}   \left[\delta\left(\lambda-\lambda_1\right)+\delta\left(\lambda-\lambda_2\right)\right]
\prod_{i=1}^{3}g(\tilde T_i) d\tilde T_i %g(\tilde T_1)g(\tilde T_2)g(\tilde T_3) %\nonumber\\ %\nonumber \\
=4 \pi \int_{0}^{\infty} r^2 dr g(r) \left[\delta\left(\lambda-\lambda_1\right)+\delta\left(\lambda-\lambda_2\right)\right],
% g(\lambda) = \int_{-\infty}^{\infty} \int_{-\infty}^{\infty}\int_{-\infty}^{\infty} g(\tilde T_1)g(\tilde T_2)g(\tilde T_3) %\nonumber\\
%  \left[\delta\left(\lambda-\lambda_1\right)+\delta\left(\lambda-\lambda_2\right)\right] d\tilde T_1 d\tilde T_2 d\tilde T_3 %\nonumber \\
% =4 \pi \int_{0}^{\infty} r^2 dr g(r) \left[\delta\left(\lambda-\lambda_1\right)+\delta\left(\lambda-\lambda_2\right)\right],
\label{eq:analytical_solution_1qubit_wn_sm}
\end{equation}
% \twocolumngrid
\end{widetext}
where the integration is performed over spherical coordinates, i.e. we substitute $r=\sqrt{\tilde T_1^2+\tilde T_2^2+\tilde T_3^2}$, and with $\delta$ denoting the Dirac delta distribution. Since 
%\begin{equation}
$g(r)=g(\tilde T_1)g(\tilde T_2)g(\tilde T_3)$ 
%\end{equation}
with $g(\tilde T_i)\propto \exp\left(-\frac{\tilde T_i^2 N}{2}\right)$, we obtain
\begin{equation}
g(r)\propto \exp\left(-\frac{r^2 N}{2}\right)
\end{equation}
with the number of counts $N$. Solving Eq.~(\ref{eq:analytical_solution_1qubit_wn_sm}), we finally obtain
\begin{equation}
g(\lambda) \propto \exp\left[-\frac{\left(1-2\lambda\right)^2 N}{2}\right] \left(1-2\lambda\right)^2.
\label{eq:definition_density_1qubit_wn}
\end{equation}
% \begin{align}
% \density(\lambda) \propto &\exp\left[-\frac{\left(1-2\lambda\right)^2 N}{2}\right] \left(1-2\lambda\right)^2 \nonumber\\
% &\left[\Theta\left(1-2\lambda\right)+\Theta\left(2\lambda-1\right)\right],
% \label{eq:definition_density_1qubit_wn}
% \end{align}
% where $\Theta\left(x\right)=1$ for $x\geq0$ ($\Theta\left(x\right)=0$ otherwise) is the Heaviside step function. 
The proportionality constant is given by normalization.

%\section{SM~5: Density for 2 and more qubits}
%\textcolor{red}{Section to be deleted!}
%The density of the eigenvalues of the $2$ qubit completely mixed state is found to be
%\begin{equation}
%\density(\lambda) \propto \exp\left(- N \alpha_2 (x-x_0)^2 - N^2 \alpha_4 (x-x_0)^4 \right) \sum_{i=0}^{3} N^i \beta_{2i} (x-x_0)^{2i}
%\end{equation}
%with $x_0=1/4$.
%The parameters  $c_0$, $c_2$, $c_4$, $c_6$, $d_2$, $d_4$ and are determined by means of fitting simulated data to the model. 
%Thereby, we obtained for $N$ counts per basis setting for our overcomplete tomography scheme
%\begin{align}
%\alpha_2(N) &\approx& 3.9\\
%\alpha_4(N) &\approx& -0.15 \; \text{PROBLEM!}\\
%\beta_0(N) &\approx& 0.0037\\
%\beta_2(N) &\approx& 0.15\\
%\beta_4(N) &\approx& -0.58\\
%\beta_6(N) &\approx& 0.75. 
%\end{align}
%Can be used as starting values...\\
%In general, the density of the completely mixed state of $n$ qubits can be described by 
%\begin{equation}
%\density(\lambda) \propto \exp\left(- N^i \sum_{i=1}^{n} \alpha_{2i} (x-x_0)^{2i} \right) \sum_{i=0}^{2^n-1} N^i \beta_{2i} (x-x_0)^{2i}.
%\end{equation}
%The damping of the density towards the borders is obtained by the leading order of the exponential term, i.e., $\exp\left(- \alpha_{2n} (x-x_0)^{2n} \right)$. 
%The power series with coefficients $\beta_{i}$ describes the comb structure.
%The completely mixed state with $n$ qubits displays a structure of $2^n$ spikes. 
%Thus, in order to describe $2^n$ spikes, a power series with leading order $2^{n+1}-2$ together with the exponential damping at the edge is needed.

\section{SM~5: Estimating the probability of physical results}
With the approximation that the eigenvalues are distributed according to a Wigner semicircle, one can estimate the probability that the result of a quantum state tomography is physical, i.e., contains only non-negative eigenvalues.
$\Tr\left(\varrho\right)=1$ holds by construction, hence, the eigenvalues are precisely speaking not independent. %, causing the last eigenvalue to be
% \begin{equation}
% \lambda_{2^n}=1-\sum_{i=1}^{2^n-1}\lambda_i,
% \label{eq:traceconstraint}
% \end{equation}
% we can estimate the probability that all eigenvalues are non-negative. % since the constraint of Eq.~(\ref{eq:traceconstraint})
However, the influence of this effect can be neglected for a sufficiently large number of qubits $n$. 
For states of the noise model given in Eq.~(\ref{eq:statedefinition}) of the main text, $2^n-1$ eigenvalues are chosen according to the respective distribution function.
% If $n^\prime$ eigenvalues 
Then, the probability that all these $\lambda_i$ are non-negative is %(for $i\leq2^n-1$) 
%\onecolumngrid
% \begin{align}
% \operatorname{Pr}\left(\lambda_i\geq0,\; i\leq2^n-1\right)=\nonumber\\
% \left[\frac{2}{\pi R^2}\int_{\max(0,c-R)}^{c+R}\sqrt{R^2-\left(\lambda-c\right)^2}\rd\lambda\right]^{2^n-1} %\equiv\operatorname{Pr}\left(\varrho\nleq0)
% \end{align}
\begin{align}
\operatorname{Pr}\left(\lambda_i\geq0\right)=\nonumber\\
\left[\frac{2}{\pi R^2}\int_{\max(0,c-R)}^{c+R}\sqrt{R^2-\left(\lambda-c\right)^2}\rd\lambda\right]^{2^n-1} %\equiv\operatorname{Pr}\left(\varrho\nleq0)
\end{align}
%\twocolumngrid
with center $c$ and radius $R$ of the semicircular distribution.
The blue line in Fig.~\ref{fig:radiusWithCounts_n6} of the main text is found by means of this calculation, being in good agreement with the simulated data, shown by points in this figure.

\section{SM~6: Different tomography schemes}
The main text focusses on the overcomplete tomography scheme.
If instead measurements are, e.g., obtained by projecting onto all $4^n$ tensor products of $|0\rangle$, $|1\rangle$, $\frac{1}{\sqrt{2}}\left(|0\rangle+|1\rangle\right)$, and $\frac{1}{\sqrt{2}}\left(|0\rangle+i|1\rangle\right)$~\cite{James}, the spectra change. 
In this tomography scheme not all correlations are measured.
One thus cannot use Eq.~(\ref{eq:Correlationsprobabilitydensity}) to describe the distribution of the correlation values.
Consequently, the Catalan numbers (see SM~2) cannot be reproduced by the moments of the distribution, leading to a different distribution than the Wigner semicircle distribution.

\begin{figure}
\includegraphics[width=0.48\textwidth]{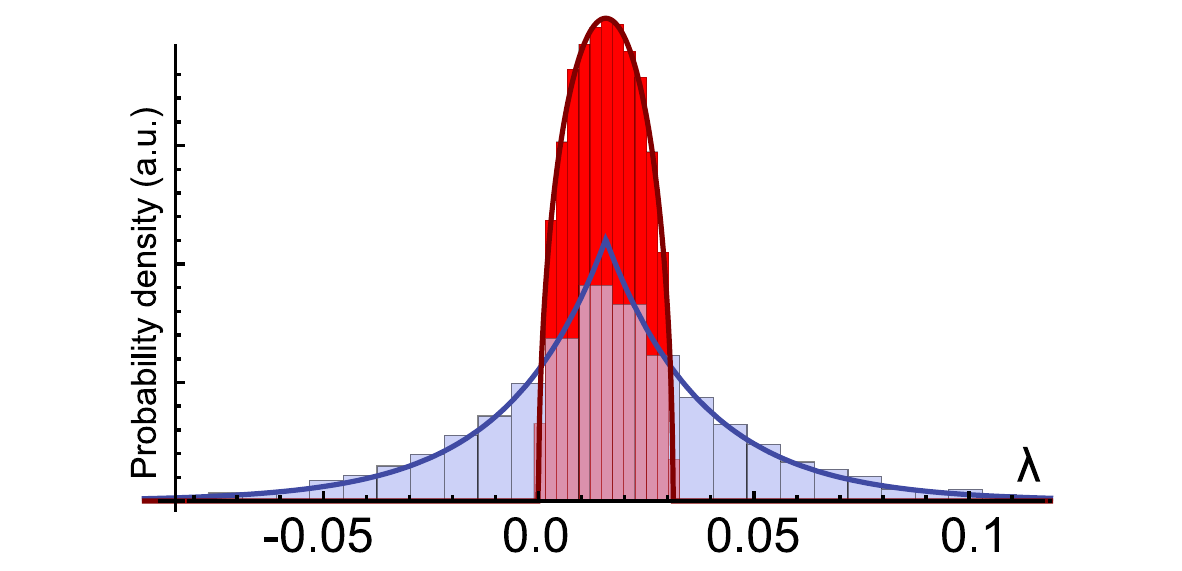}
\caption{Simulated spectrum of $6$ qubit completely mixed state obtained using the tomographically complete scheme~\cite{James} (blue bars) and using the overcomplete scheme as used in the main text (red bars). 
The blue line is given by $\operatorname{h}\left(\lambda\right)$ according to Eq.~(\ref{eq:expdistributionParams}).
The red line corresponds to the semicircle model derived in the main text. 
While all eigenvalues in the latter scheme are expected to be positive, the asymmetric scheme~\cite{James} gives unphysical results in all cases.}
\label{fig:eigsCompleten6N62500}
\end{figure}

Fig.~\ref{fig:eigsCompleten6N62500} compares complete and overcomplete sampling on a basis of the spectral distribution for $100$ simulated states ($n=6$ qubits, completely mixed state, cf. Fig.~\ref{fig:wn_n6_N100} of the main text).
We choose the total number of measurements $N_{\rm total}=4\,000\,000$ since, according to Eq.~(\ref{eq:effectiveRadius}), when distributing all measurements over the required $3^6$ settings ($N=N_{\rm total}/3^6\approx5487$), one obtains physical estimates for overcomplete sampling with the eigenvalue distribution shown in red.
On the contrary, for complete sampling~\cite{James} one obtains the distribution shown in blue. 
The blue line corresponds to $\operatorname{h}\left(\lambda\right)=\frac{\alpha}{2}\exp\left(-\alpha\left|\lambda-2^{-n}\right|\right)$,
where $\alpha$ now can be estimated by the method of moments (as used in SM~2). 
The second (centralized) moment of the distribution $\operatorname{h}\left(\lambda\right)$ is found to be $2/\alpha^2$ and has to be equal to the second (centralized) moment of the eigenvalue distribution, i.e., $4^n/N_{\rm total}$, where $N_{\rm total}$ denotes the total number of measurement events. % \textcolor{red}{$\leftarrow$ check this!}. %given in Eq.~(\ref{eq:expdistribution}) 
%For the case of $n=6$ qubits, we estimate $\alpha\approx1250/\sqrt{N_{\rm total}}$.
Thus, the spectral probability distribution for the informationally complete tomography as proposed in Ref.~\cite{James} can be approximated by
\begin{equation}
\operatorname{h}\left(\lambda\right)=\sqrt{\frac{N_{\rm total}}{2\cdot 4^n}}\exp\left(-\sqrt{\frac{2 N_{\rm total}}{4^n}}\left|\lambda-2^{-n}\right|\right).
\label{eq:expdistributionParams}
\end{equation}
Hence, this tomography scheme reports eigenvalues with high probability in the tails of the distribution far away from the center at $1/64$.
For the dark blue line in Fig.~\ref{fig:eigsCompleten6N62500}, one obtains $\alpha=\sqrt{2\cdot4\cdot10^6/4^6}\approx44.2$.

\section{SM~7: Modification of radius for low rank states}
\begin{figure}
\includegraphics[width=0.48\textwidth]{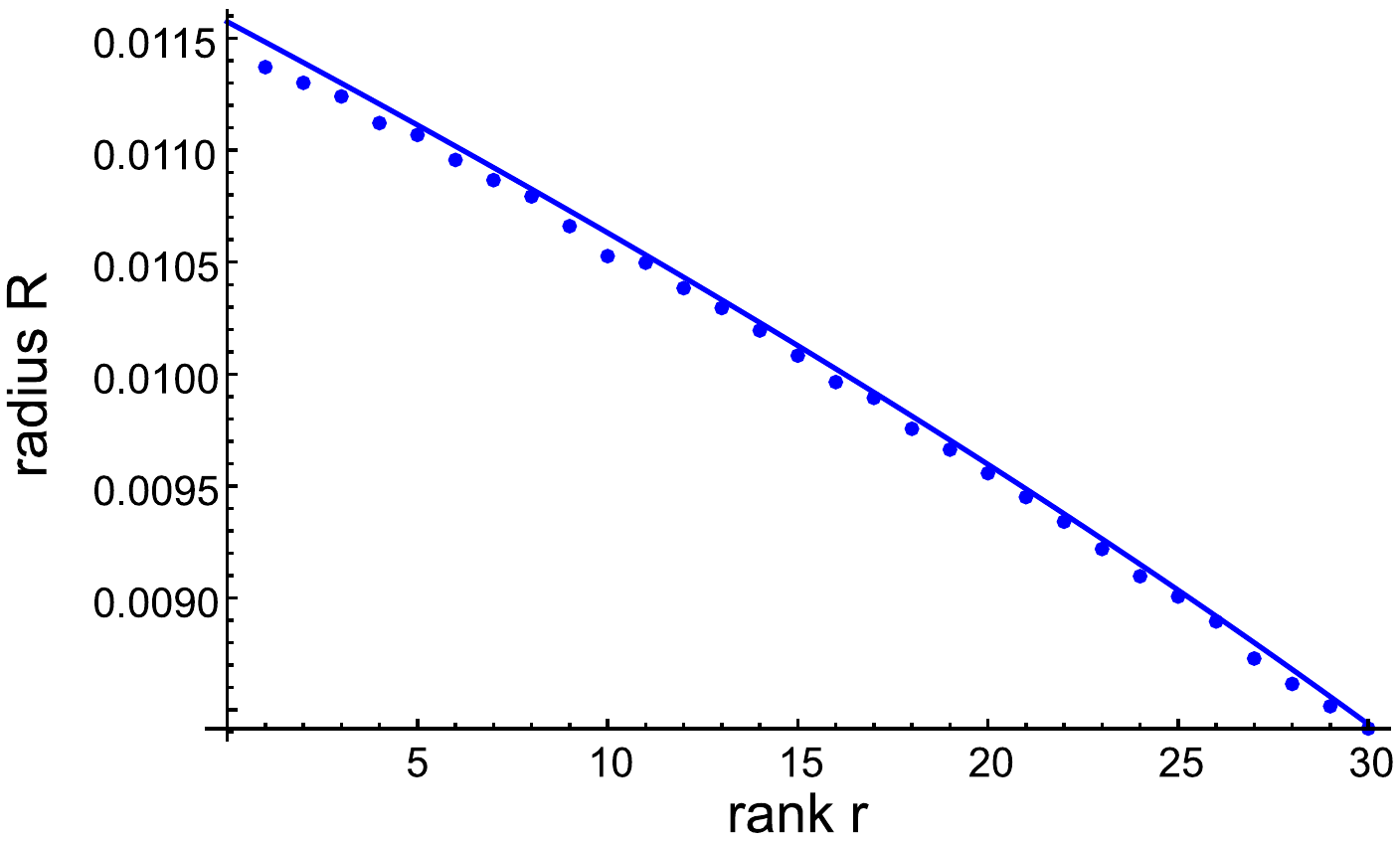}
\caption{QST (with $N=10\,000$) is simulated for the equal mixture of $r$ pure (random) $n=6$ qubit states. 
The numerically obtained radii $R$ for different values for the rank $r$ are compared with the expected radius behaviour of Eq.~(\ref{eq:modifiedradius}).
}
\label{fig:radiusWithRank_n6_N10000}
\end{figure}

According to SM~2, the spectral probability density of the maximally mixed state converges to the Wigner semicircle. 
For low rank states, the radius of the spectral probability distribution of the noise can be approximated with the radius as given in Eq.~(\ref{eq:effectiveRadius}).
The assumption that the radius does not depend on the rank $r$ is only an approximation valid for $r\ll2^n$.
In Fig.~\ref{fig:radiusWithRank_n6_N10000}, the radius $R_r$ in dependence of the rank $r$ is shown for $100$ QST estimates ($N=10\,000$), where the underlying state is obtained by equally mixing $r$ orthogonal random states.
$R$ can be estimated by means of the second moment $m^{\rm ev}_2=1/2^n \E\left[\sum_i\lambda_i^2\right]=\left(R/2\right)^2$.
For a low rank state, the summation over the eigenvalues in Eq.~(\ref{eq:paulidirac}) has to be modified, 
\begin{align}
m^{\rm ev}_2\rightarrow m^{\rm ev\prime}_2\approx m^{\rm ev}_2 \cdot \left(\frac{2^n - r}{2^n}\right)=m^{\rm ev}_2 \cdot \left(1-r \cdot 2^{-n}\right).\\
R\rightarrow R^\prime_r=R \cdot \sqrt{1-r \cdot 2^{-n}} = 2\left(\frac{5}{6}\right)^{\frac{n}{2}}\frac{\sqrt{1-r \cdot 2^{-n}}}{\sqrt{N}}. 
\label{eq:modifiedradius}
\end{align}
%Thus, the radius this modification can be taken into account when evaluating the radius,
%With this modification, one obtains an estimate for the radius,
% Consequently, the radius is modified as
% \begin{equation}
% % R\left(n,N\right)\rightarrow R^\prime\left(n,N,r\right)=R\left(n,N\right) \cdot \sqrt{1-r \cdot 2^{-n}} = 2\left(\frac{5}{6}\right)^{\frac{n}{2}}\frac{1}{\sqrt{N}} \cdot \sqrt{1-r \cdot 2^{-n}}.
% R\rightarrow R^\prime_r=R \cdot \sqrt{1-r \cdot 2^{-n}} = 2\left(\frac{5}{6}\right)^{\frac{n}{2}}\frac{1}{\sqrt{N}} \cdot \sqrt{1-r \cdot 2^{-n}}. 
% \label{eq:modifiedradius}
% \end{equation}
For a rank $r=1$ state with $n=6$ qubits, the radius is changed by a factor of $\sqrt{1-r\cdot2^{-n}}\approx0.992$, which can be neglected.

\section{SM~8: Hypothesis test}
In the main text, the obtained spectrum of the experimentally prepared Dicke state~\cite{Dicke63} is tested against its anticipated distribution. 
For this hypothesis test, we use the Anderson-Darling test~\cite{AndersonDarling}. 
The distance between the hypothetical cumulative distribution function (CDF) $F\left(x\right)$ and the empirical distribution function (EDF) $\tilde{F}\left(\lambda\right)$ is calculated by means of
\begin{equation}
{\cal D}=\overline{n}\int_{-\infty}^{\infty}{w\left(\lambda\right)\left[\tilde{F}\left(\lambda\right)-F\left(\lambda\right)\right]^2\rd F\left(\lambda\right)},
\end{equation}
where $w\left(\lambda\right)$ is a weighting function and where $\overline{n}$ denotes the sample size (number of eigenvalues in empirical data).

The function $w\left(\lambda\right)=1/\left[F\left(\lambda\right)\left[1-F\left(\lambda\right)\right]\right]$ used for the Anderson-Darling test~\cite{AndersonDarling} weights the tails of the distribution higher than, e.g., the Cram{\'e}r-von-Mises test~\cite{CramerMises}.
The CDF of a semicircle with radius $R$ and center $c$ reads
\begin{equation}
F\left(\lambda\right)=\frac{1}{2}+\frac{\left(\lambda-c\right)\sqrt{1-\frac{\left(\lambda-c\right)^2}{R^2}}}{\pi R}+\frac{\operatorname{ArcSin}\left(\frac{\left(\lambda-c\right)}{R}\right)}{\pi}
\end{equation}
(for $c-R<\lambda<c+R$; $F\left(\lambda\right)=0$ for $\lambda\leq c-R$ and $F\left(\lambda\right)=1$ for $\lambda\geq c+R$).
% \onecolumngrid
% \begin{equation}
% F\left(\lambda\right)=\begin{cases} \frac{1}{2}+\frac{\left(\lambda-c\right)\sqrt{1-\frac{\left(\lambda-c\right)^2}{R^2}}}{\pi R}+\frac{\operatorname{ArcSin}\left(\frac{\left(\lambda-c\right)}{R}\right)}{\pi} &\mbox{if } c-R<\lambda<c+R \\
% 0 & \mbox{if } \lambda\leq c-R\\
% 1 & \mbox{if } \lambda\geq c+R\end{cases} 
% \end{equation}
% \twocolumngrid
%  and the EDF $\tilde{F}\left(\lambda\right)$, the Anderson-Darling test~\cite{AndersonDarling} reads
% \begin{equation}
% A=\overline{n}\int_{-\infty}^{\infty}{\frac{\left[\tilde{F}\left(\lambda\right)-F\left(\lambda\right)\right]^2}{F\left(\lambda\right)\left[1-F\left(\lambda\right)\right]}\rd F\left(\lambda\right)}.
% \end{equation}
Please note that the difference of the empirical and the cumulative distribution function contributes only for $\rd F\left(\lambda\right)\neq0$ to the test statistic from which the $P$-value is derived.
Thus, if an eigenvalue is outside of the support of the Wigner semicircle distribution, it does not matter how far it is off.
Consequently, it is recommended to check in the first place that all the data are in the support and set the $P$-value to $0$ otherwise.
\begin{figure}
\includegraphics[width=0.42\textwidth]{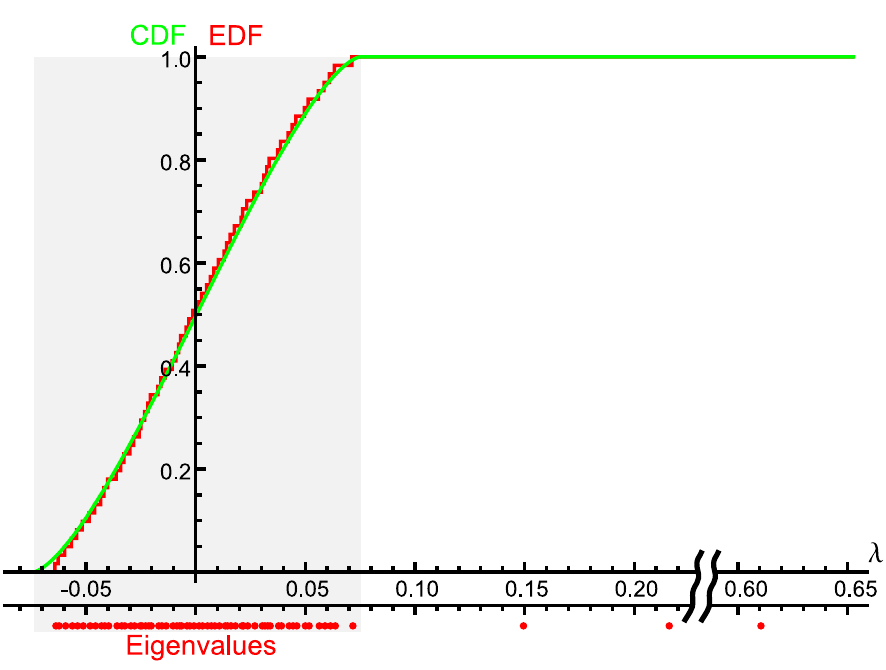}
\caption{(red) Empirical distribution function (EDF) of the smallest $61$ eigenvalues, i.e., it is given by $|\{\lambda_i:\lambda_i\leq \lambda, i\leq 61\}| / 61$, where $|S|$ denotes the number of elements in set $S$.
(green) Cumulative distribution function (CDF) of the Wigner semicircle distribution with $R=0.0745$ and $c=4\times 10^{-4}$, in good agreement with the red EDF.
The eigenvalues are shown as red dots underneath the plot.
The support of the assumed Wigner semicircle distribution is shown by the gray shaded area.
}
\label{fig:dicke63_eigenvalues}
\end{figure}

\begin{center}
\begin{table}[t]
\caption{Center and radius of a hypothetical Wigner semicircle distribution %and the corresponding $P$-Values 
for different ranks $r$.
The $P$-values directly obtained by means of an Anderson-Darling (AD) hypothesis test are shown as well as \textit{effective} $P_{\rm eff}$-values which account for eigenvalues outside of the support of the semicircle.}
\vspace{0.2cm}
\begin{tabular}{| r | r | r | r | r | } 
\hline \hline 
rank $r$ \hspace{0.0cm}& \hspace{0.0cm}center $c$ 	\hspace{0.0cm}& \hspace{0.0cm}radius $R$ \hspace{0.0cm} & \hspace{0.0cm} $P$-value \hspace{0.0cm} & \hspace{0.0cm}$P_{\rm eff}$-value \hspace{0.0cm}\\ \hline
$0$ 	&  $0.015625$ & $0.076317$ & $9.44\cdot 10^{-6}$& $0$ \\ \hline
$1$ 	&  $0.006187$ & $0.075719$& $0.0089$ & $0$   \\ \hline
$2$ 	&  $0.002803$ & $0.075115$& $0.3553$ & $0$  \\ \hline
$3$ 	&  $0.000399$ & $0.074507$& $1-8\cdot10^{-7}$ & $1-8\cdot10^{-7}$ \\ \hline
$4$ 	& $-0.000790$ & $0.073894$& $0.9998$ & $0.9998$ \\ \hline
$5$ 	& $-0.001883$ & $0.073275$& $0.9976$ & $0.9976$ \\ %\hline
%$6$	& $-0.002976$ & $0.072652$&	$0.802$ \\
\hline \hline
\end{tabular}
\label{tab:pVals}
\end{table}
\end{center}

As an additional check of the state estimation strategy introduced above, one can compare the $P$-values for various ranks $r$ assumed in the tests.
In Tab.~\ref{tab:pVals} the center $c$ and the corresponding $P$-values are given for low rank states with admixed white noise for different ranks $r$.
Please note that for the cases where eigenvalues are found outside of the support of the Wigner semicircle distribution the corresponding hypothesis that statistical noise can be an explanation for these eigenvalues has to be rejected, effectively leading to a vanishing $P$-value.
 
Evidently, for $r\geq4$, the center of the assumed semicircle is shifted to negative values, being incompatible with our model. 
Consequently, the analysis results in a rank $r=3$ state with a small amount %($1-q\approx0.024$) 
of admixed white noise.

\end{document}